\documentclass[a4paper,11pt]{article}
\pdfoutput=1 

\usepackage{jcappub} 

\usepackage[T1]{fontenc} 
\usepackage{graphicx}	
\usepackage{subfig}
\usepackage{amsmath}	
\usepackage{empheq}
\usepackage{amssymb}
\usepackage{natbib}
\newcommand{\dm}{\delta_{\rm{m}}}
\newcommand{\dnu}{\delta_{\nu}}
\newcommand{\fnu}{f_{\nu}}

\newcommand{\dcb}{\delta_{\rm{cb}}}
\newcommand{\fone}{{}_2{\rm{F}}_1}
\newcommand{\aT}{a_{\rm{T}}}

\title{\boldmath A Simple Analytic Treatment of Neutrino Mass Impact on the Full Power Spectrum Shape via a Two-Fluid Approximation}

\author[a,1]{Farshad Kamalinejad\note{Corresponding author.}} 
\author[b]{\& Zachary Slepian}
\emailAdd{f.kamalinejad@ufl.edu}
\emailAdd{zslepian@ufl.edu}

\affiliation[a]{Department of Physics, University of Florida, 2001 Museum Road, Gainesville, FL 32611, USA}
\affiliation[b]{Department of Astronomy, University of Florida, 211 Bryant Space Science Center, Gainesville, FL 32611, USA}

\abstract{We present a new closed-form formula for the matter power spectrum in the presence of massive neutrinos that gives an accuracy of better than 5\% on all scales. It is the first closed-form result valid on all scales. To calculate this closed-form solution, we iteratively solve the fluid equations for cold dark matter+baryons, and neutrinos in terms of the neutrino mass fraction, $f_{\nu} \ll 1$, using variation of parameters to construct the response of the matter to the neutrino perturbation, which in turn will source the higher-order neutrino perturbations. This analytic solution accelerates calculations of the matter power spectrum with neutrinos. Also, it enables one to calculate the cross power spectra of matter and neutrinos which in turn can be used in computing the  higher-order corrections to the power spectrum. To demonstrate our formula's accuracy and utility, we perform a Fisher forecast with it and show we reproduce the forecast from the Boltzmann solver \textsc{class}.} 

\begin{document}
\maketitle
\flushbottom

\section{Introduction}

In the Standard Model (SM) of particle physics, neutrinos are massless, but in fact, they have non-zero mass. This already indicates that there is physics beyond the SM, making further investigation into neutrino mass one of the most exciting areas at the intersection of high-energy physics and cosmology.

Before we discuss why cosmology in particular has a bearing on the neutrino mass, we first briefly review how we know neutrinos have mass. Neutrinos exist in three flavors: electron ($\nu_{e}$), muon ($\nu_{\mu}$), and tau ($\nu_{\tau}$), and these flavors are each linear combinations of mass eigenstates. Neutrinos are produced in flavor eigenstates but propagate in mass eigenstates, meaning that a beam of neutrinos arrives with different proportions of the flavors than those with which it was initially produced. This process is known as neutrino oscillation, a phenomenon predicted by \cite{Pontecorvo:1957cp, Pontecorvo:1957qd}. However, it is only sensitive to the differences in the squares of the mass states, which means it can constrain only the relative, not the absolute, mass scale. The most recent measurements from neutrino oscillation experiments \cite{2004, abe2021t2k, abe2024first, esteban2020fate} reveal that:
\begin{align}
    \Delta m_{21}^2 = \left(7.42^{+0.21}_{-0.20} \right)\times 10^{-5}\; {\rm{eV}}^2, \;\;\vert\Delta m_{31}^2 \vert=\left(2.517^{+0.026}_{-0.028} \right)\times 10^{-3}\; \rm{eV}^2.
\end{align}
Determining the absolute mass scale, which would be possible with one additional constraint, such as the sum of the masses, is the next step in understanding how neutrinos guide us to what lies beyond the SM. Most immediately, measuring the absolute mass scale would reveal whether the mass hierarchy is normal (two heavy, one light) or inverted (two light, one heavy). The neutrino mass ordering also determines whether neutrinos are Dirac or Majorana particles and will enable better understanding of the theory of neutrino flavors.  \cite{Qian_2015, cahn2013whitepapermeasuringneutrino}.

To measure the absolute mass scale, terrestrial experiments such as KATRIN use tritium beta decay, but it is very challenging to reach high enough sensitivity. The current bound on the electron neutrino mass scale from KATRIN is $m_{\nu_e}<0.8\,{\rm eV}$, which implies the total mass scale for three degenerate mass states is $\sum m_{\nu}<0.24\,{\rm eV}$ \cite{adame2024desi}. KATRIN ultimate precision by its conclusion is predicted to be $m_{\nu_e}<0.2\,{\rm eV}$ \cite{aker2022katrin}.

Thus, we need another probe of the neutrino mass sum. The Universe can be regarded as a laboratory for neutrino physics, since a cosmological background of relic neutrinos is predicted in the standard hot Big Bang paradigm, $\rm{\Lambda CDM}$ (cosmological constant $\Lambda$ and Cold Dark Matter). According to $\rm{\Lambda CDM}$, the neutrinos decoupled from the rest of the matter when the weak force interactions became insufficient in comparison to the expansion of the Universe, when the temperature was about $T\approx 1\;\rm{MeV}$. Neutrinos then affect the Comic Microwave Background (CMB) radiation. The $95\%$ upper bound on the neutrino mass from the CMB temperature and polarization measurements \cite{aghanim2020planck} is $\sum m_{\nu}<0.26\;{\rm eV}$. Combining the CMB and large-scale structure measurements will tighten the bound to $\sum m_{\nu}<0.12\,{\rm eV}$ to which latter we now turn.

In addition to the effects on the CMB, the neutrinos also affect the large-scale structure due their free-streaming. The expansion of the Universe dilutes the momentum of neutrinos, leading them to eventually become non-relativistic. Once the transition occurs, the neutrinos cluster on large scales. On smaller scales, however, they do not cluster due to their high thermal velocities; this effect suppresses the growth of structure on these scales. The characteristic length scale separating the large-scale and small-scale behavior is called the free-streaming scale. The suppressed growth affects galaxy clustering and can be probed by current and upcoming galaxy surveys such as the extended Baryon Oscillation Spectroscopic Survey (eBOSS) \cite{alam2021completed}, Dark Energy Spectroscopic Instrument (\textsc{DESI}) \cite{DESI:2016, adame2024desi}, \textsc{Euclid} \cite{euclid2021euclid}, and \textsc{Roman}  Space Telescope \cite{wang2022high}.


In the context of linear cosmological perturbation theory, the neutrinos affect the matter power spectrum, leading to a constant suppression on scales smaller than $10 \;h {\rm Mpc}^{-1}$ \cite{Hu_1998}. On larger scales (between $10 \;h {\rm Mpc}^{-1}$ and $100 \;h {\rm Mpc}^{-1}$) the neutrinos reduce the power spectrum, in a scale-dependent manner. Modeling this reduction requires running Boltzmann solvers such as \textsc{camb} \cite{lewis2011camb} or \textsc{class} \cite{2011jul} to compute the full power spectrum numerically.



Although highly accurate and precise predictions from Boltzmann codes exist, a simple analytic  treatment of the shape of the matter power spectrum on intermediate regimes seems desirable. This analytic  formula could improve understanding and potentially accelerate the use of the power spectrum with neutrino mass.

In this paper, our main focus is to analytically solve the evolution equations (two-fluid equations: neutrinos and matter \cite{shoji2010massive}) in the linear regime. Our goal is to present a simple and understandable picture where neutrinos are treated so as to enable us to predict analytically the shape of the power spectrum across all scales. Our approach is to expand the two-fluid equations in the neutrino mass fraction, $\fnu$, which is small, and then solve for the effects of neutrinos on the matter density perturbation iteratively. Our result gives an accurate closed-form formula (which is much faster than running a Boltzmann code) that can be used in Fisher forecasting, obtaining an approximation to the cross power spectra of cold dark matter and of baryons with neutrinos, which is necessary in calculating the higher-order corrections to the matter power spectrum \cite{Wong_2008} with the FFTlog method \cite{mcewen2016fast, schmittfull2016fast, simonovic2018cosmological}.

As noted above, the neutrinos cause a constant suppression on small scales in linear theory. This constant suppression can only be seen on scales smaller than $10\;{\rm Mpc}/h$ (in Fourier space, $k>0.1\;h {\rm Mpc}^{-1})$. However, linear perturbation theory breaks down on these scales. To understand the neutrino effects on these scales, one must therefore go beyond linear theory. The first attempts to include massive neutrinos in non-linear structure formation perturbatively were \cite{Saito_2008} and \cite{Wong_2008}. They assumed the neutrinos are only perturbed up to the linear order, neglecting the higher-order contributions. This approximation is discussed in \cite{Blas_2014} and as it turns out, ignoring the small-scale neutrino perturbations results in violation of momentum conservation. \cite{Wong_2008, Saito_2008} also numerically solved the two-fluid equations for non-linear structure formation in the  presence of massive neutrinos. \cite{2021} also studies the clustering of massive neutrinos in Eulerian perturbation theory and shows promising agreement with simulations on small scales. Although our work gives accurate results for scales smaller than $10\;{\rm Mpc}/h$, we are mostly interested in the regime where linear perturbation theory still applies (scales $\gtrsim 10\;{\rm Mpc}/h$).


The structure of this paper is as follows. In \S\ref{Neutrino Properties} we discuss the basic neutrino properties, then in \S\ref{Boltzmann Equation} we obtain the fluid-equations and discuss the free-streaming scale. In \S\ref{Two-fluid equations} we discuss the the two-fluid system and in \S\ref{Iterative Method} we analytically find the closed-form power spectrum with massive neutrinos. Then we obtain the asymptotic behavior of our model in \S\ref{limiting form} and show that our model recovers the previous results in the literature. In \S\ref{Fisher Forecast}, we test our analytic model in a Fisher forecast against the predictions of the \textsc{class} code and show how well our iterative method compares with \textsc{class} for this forecast. Finally, we will also comment on the impact of dark energy on the power spectrum suppression in \S\ref{Dark Energy Modifications}. Throughout this paper, we use the \texttt{python} package \textsc{Nbodykit} \cite{Hand_2018} which relies on \textsc{class} as an engine for the underlying cosmological calculations \cite{2011jul}.

\section{Neutrino Dynamics}
\label{Neutrino Properties}

\subsection{Neutrino Equation of State, Energy Density, and Transition Time}
The neutrinos decoupled from the primordial plasma when $T \approx 1\; \rm{MeV}$ and subsequently free-streamed. If the total comoving energy of the $i^{th}$ neutrino mass state is $\epsilon_i$, we have that $\epsilon_i = \sqrt{q_i^2+a^2m_{\nu,i}^2}$ where $q_i$ is the comoving momentum of the $i^{th}$ state and $m_{\nu,i}$ its rest mass \cite{shoji2010massive}. As the Universe expands, the momentum of the neutrinos decays as $1/a$; once their energy is dominated by the rest mass, the neutrinos become non-relativistic. This transition occurs when the neutrinos' average momentum falls below the rest mass. The average momentum of the neutrinos when they are ultra-relativistic is obtained from the Fermi-Dirac distribution function as \cite{2014,Levi16}:
\begin{align}
    \left<p\right> &= \frac{\displaystyle\int d^3\vec{p}\; p \left(e^{p/T_{\nu}} + 1\right)^{-1}}{\displaystyle\int d^3\vec{p}\; \left(e^{p/T_{\nu}} + 1\right)^{-1}} = 3.15 \; T_{\nu,0}(1+z),
\label{mean momentum}
\end{align}
where $T_{\nu,0} = \left(4/11 \right)^{1/3}T_{\rm CMB,0}$ is the present-time neutrino temperature and $T_{\rm CMB,0}$ is the CMB temperature. Therefore the transition redshift for the $i^{th}$ mass state, $z_{\rm{T},i}$, can be written as \cite{shoji2010massive}:
\begin{align}
    1+z_{\rm{T},i} = 1890 \left(\frac{m_{\nu,i}}{1\;\rm{eV}}\right).
\label{transition redshift}
\end{align}
Assuming the neutrino masses are below $1\;\rm{eV}$, the transition occurs when the Universe is already matter dominated \cite{LESGOURGUES_2006}.

We recall that the transition redshift is when the kinetic energy of particles is equal to their rest mass energy; consequently, the neutrinos are not completely non-relativistic at the time of transition. As an approximation, we take it that the neutrinos' momentum distribution function is a Dirac delta function centered at $\left<p\right>$, given by (\ref{mean momentum}). Following the treatment of \cite{Slepian_2018_neutrino,1981SvA....25..521D}, which give this approximation, we find the equation of state and density of the neutrinos as:
\begin{align}
    w(a) &= \frac{1}{3}\left(\frac{1}{1+(a/a_{\rm T})^2}\right),
    \label{EOSparameter}
\end{align}
and
\begin{align}
    \rho_{\nu}(a) &= \rho_{\nu,0}a^{-3} g^{1/2}(a),
    \label{neutrino density}
\end{align}
with 
\begin{align}
    g(a) = \frac{1+(a_{\rm T}/a)^2}{1+a_{\rm T}^2}
    \label{g(a)}.
\end{align}
The transition scale factor is approximately given by $a_{\rm T}\approx1/z_{\rm T}$ since $z_{\rm T}\gg1$. $\rho_{\nu,0}$ is the present-time neutrino density. Inspecting the limits of (\ref{g(a)}) suggests that at early times, when $a\ll a_{\rm T}$, the neutrinos behave essentially as radiation. At late times however, $g(a)\rightarrow1$, and therefore the neutrinos evolve as matter.

The neutrino mass fraction is defined as the ratio of the neutrino density parameter to the total matter density:\footnote{We distinguish between $\fnu(\tau)$ and $\fnu$, which latter is the present-time neutrino fraction. The time dependence will be made explicit anywhere we do not mean the present-time neutrino fraction.}
\begin{subequations}
\begin{align}
    \fnu(a) &\equiv \frac{\Omega_{\nu}(a)}{\Omega_{\rm m}(a)}=\fnu\; g^{1/2}(a)\label{fnu}\\
    &\approx \fnu\left(1+\frac{1}{2}\left(\frac{a_{\rm T}}{a}\right)^2+\cdots\right)
    \label{fnuII},
\end{align}
\end{subequations}
where 
\begin{align}
   \fnu=\frac{1}{\Omega_{{\rm m}}h^2}\;\frac{\sum_{i} m_{\nu,i}}{93.14 \,\rm{eV}}
    \label{fnu0}
\end{align}
is the present-time neutrino mass fraction. $h$ is the Hubble constant in units of $100\; {\rm km}\, {\rm s}^{-1}{\rm Mpc}^{-1}$, and $\Omega_{\rm m}$ is the present-time total matter density in units of the critical density, $3H_0^2 / (8 \pi G)$, with $H_0$ the value of the Hubble parameter at present and $G$ Newton's constant. We have explicitly emphasized the time-dependence of the variables wherever this is relevant. Otherwise, we refer to their present-time value.
\begin{figure}
    \centering
    \includegraphics[width=7cm]{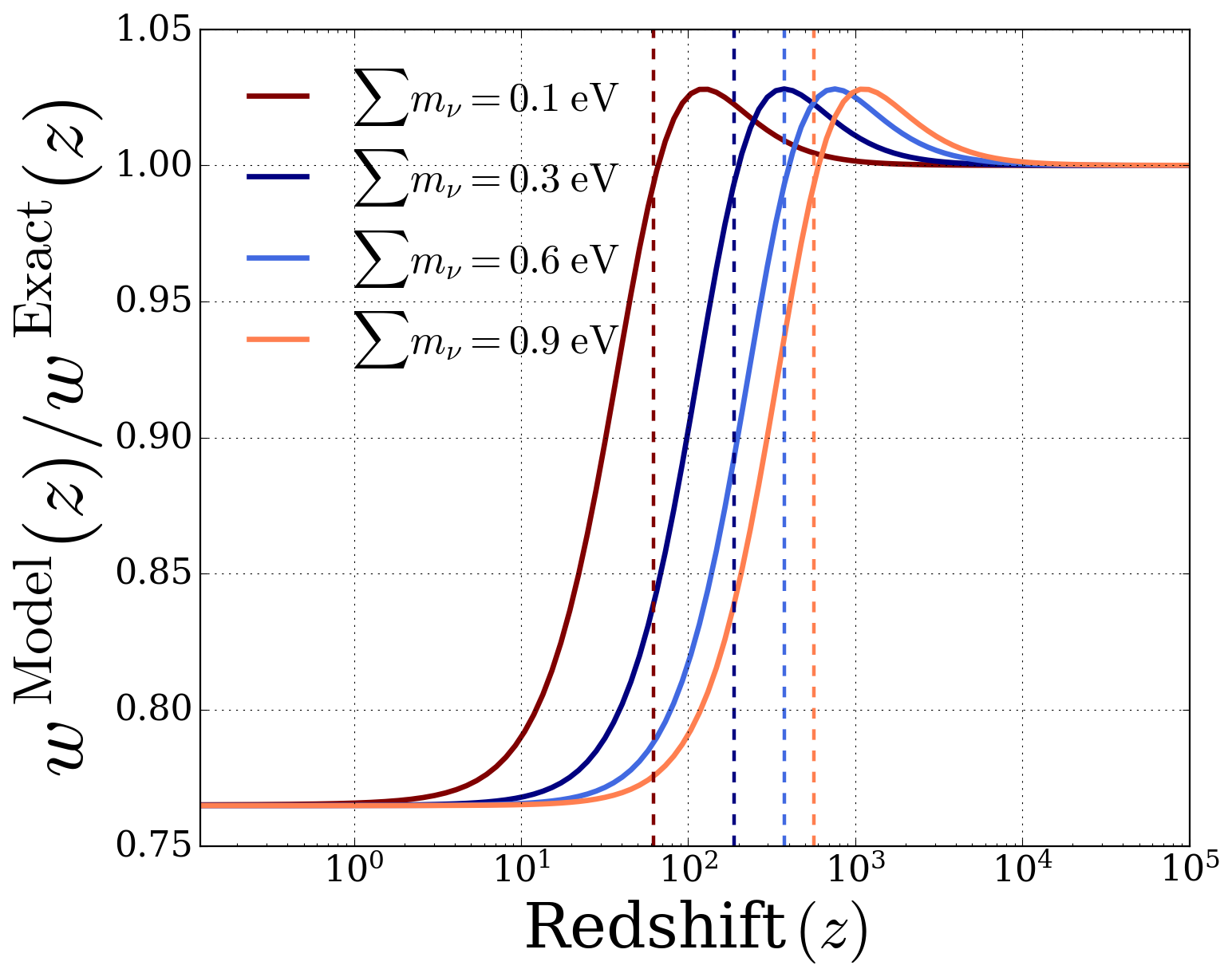}
    \qquad
    \includegraphics[width=7cm]{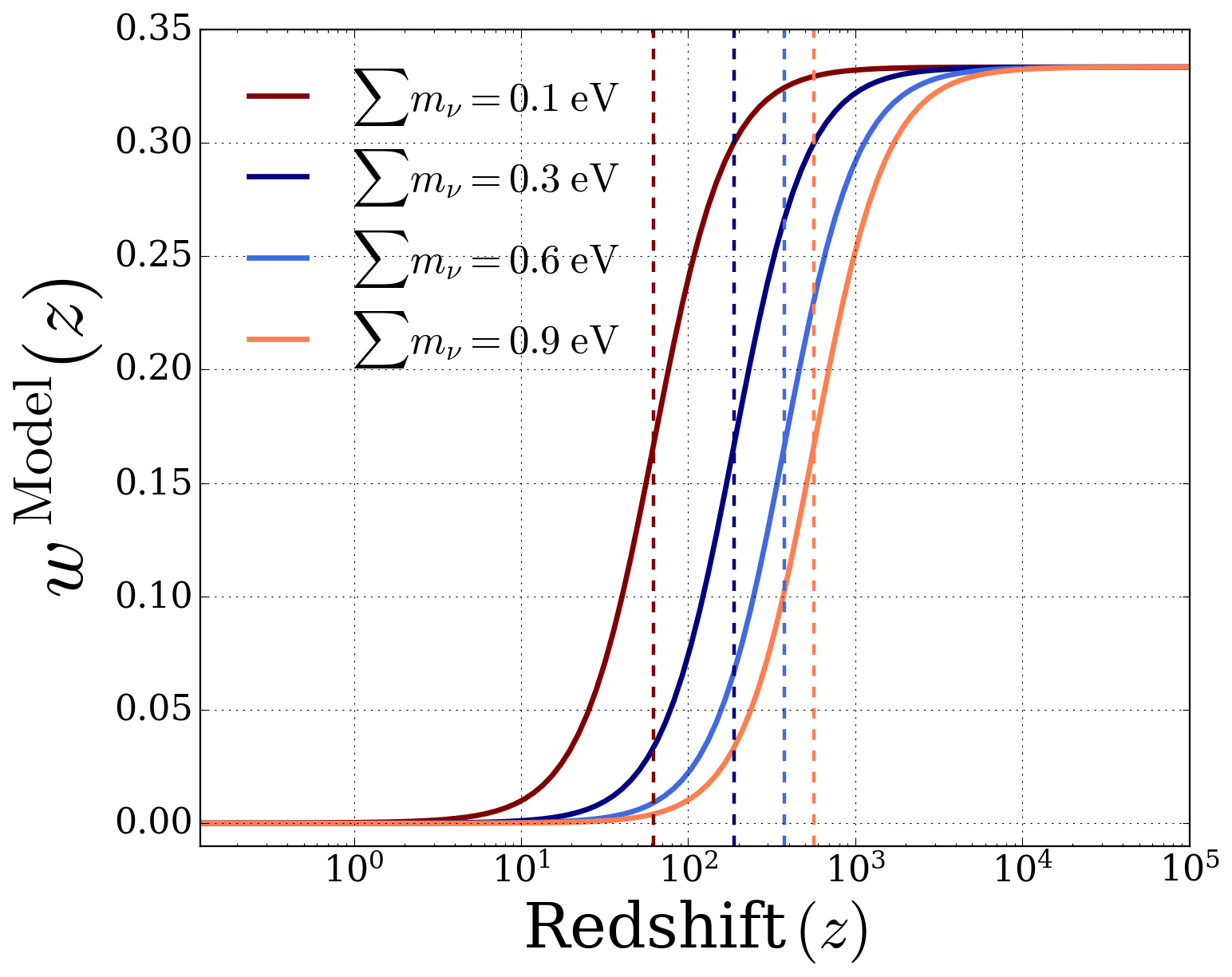}%
    \caption{\textit{Left:} The ratio of our neutrino equation of state to that from \textsc{class}. The current best bound is $\sum m_{\nu} = 0.26\; {\rm eV}$, from the CMB \cite{aghanim2020planck}. This panel shows that our approximation is accurate to $80\%$ at late times for a broad range of neutrino masses. Around the transition redshift, the small bumps are due to the approximation being imperfect when the momentum is comparable to the neutrino mass. $w(z)$ enters the two-fluid equations that we will soon derive solely through the free-streaming scale \S\ref{Boltzmann Equation}, not the evolution equations; importantly, the low-redshift disagreement of our approach with \textsc{class} changes the free-streaming scale by a negligible amount. \textit{Right:} Our model equation of state for the neutrinos (\ref{EOSparameter})  vs. redshift for a range of neutrino masses. The transition redshift (where the e.o.s. falls from roughly $1/3$ down to zero) increases with the neutrino mass. The dashed vertical lines are the transition redshift corresponding to each neutrino mass.}
    \label{w(z)}
\end{figure}

\begin{figure}
    \centering
    \includegraphics[width=7cm]{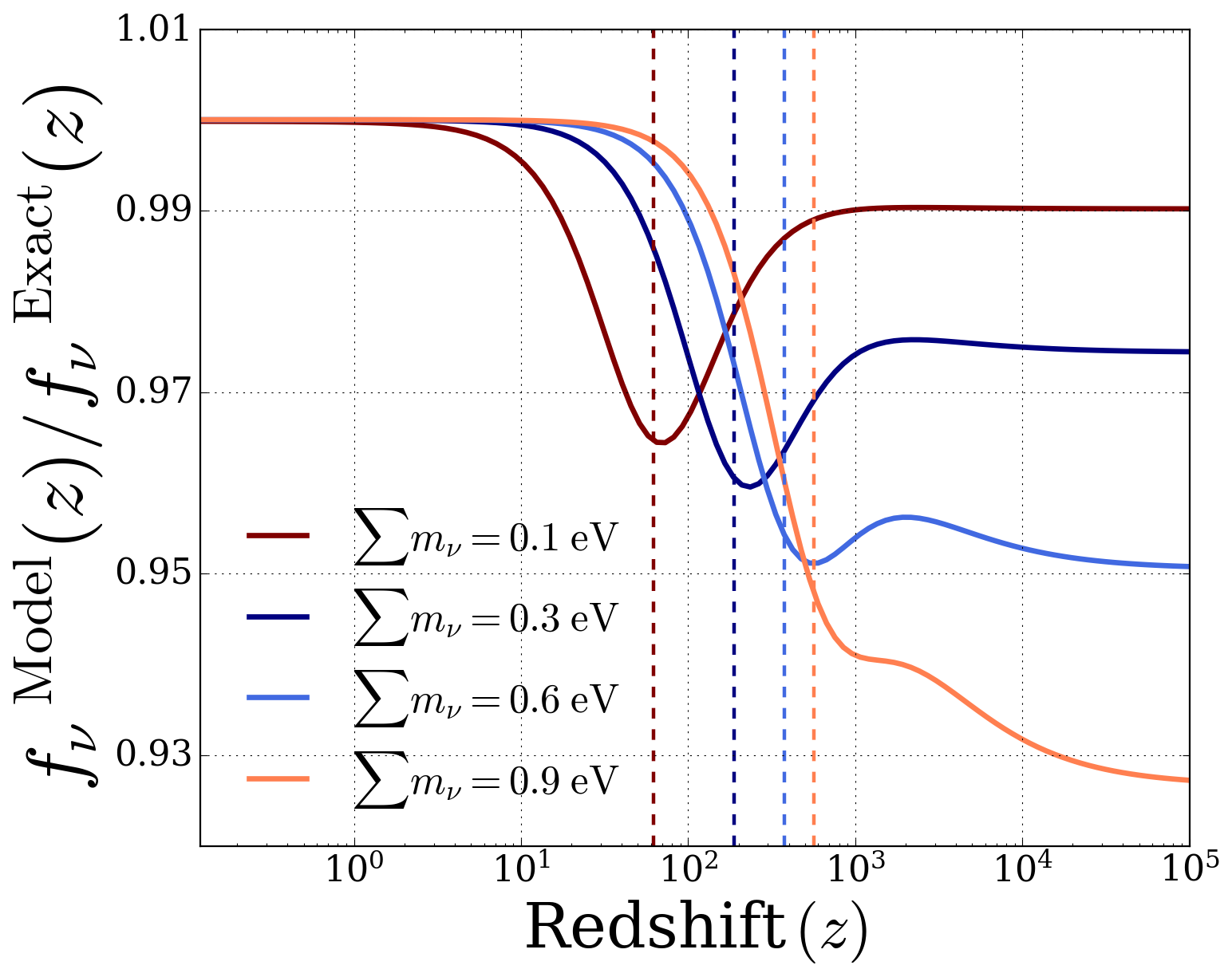}
    \qquad
    \includegraphics[width=7cm]{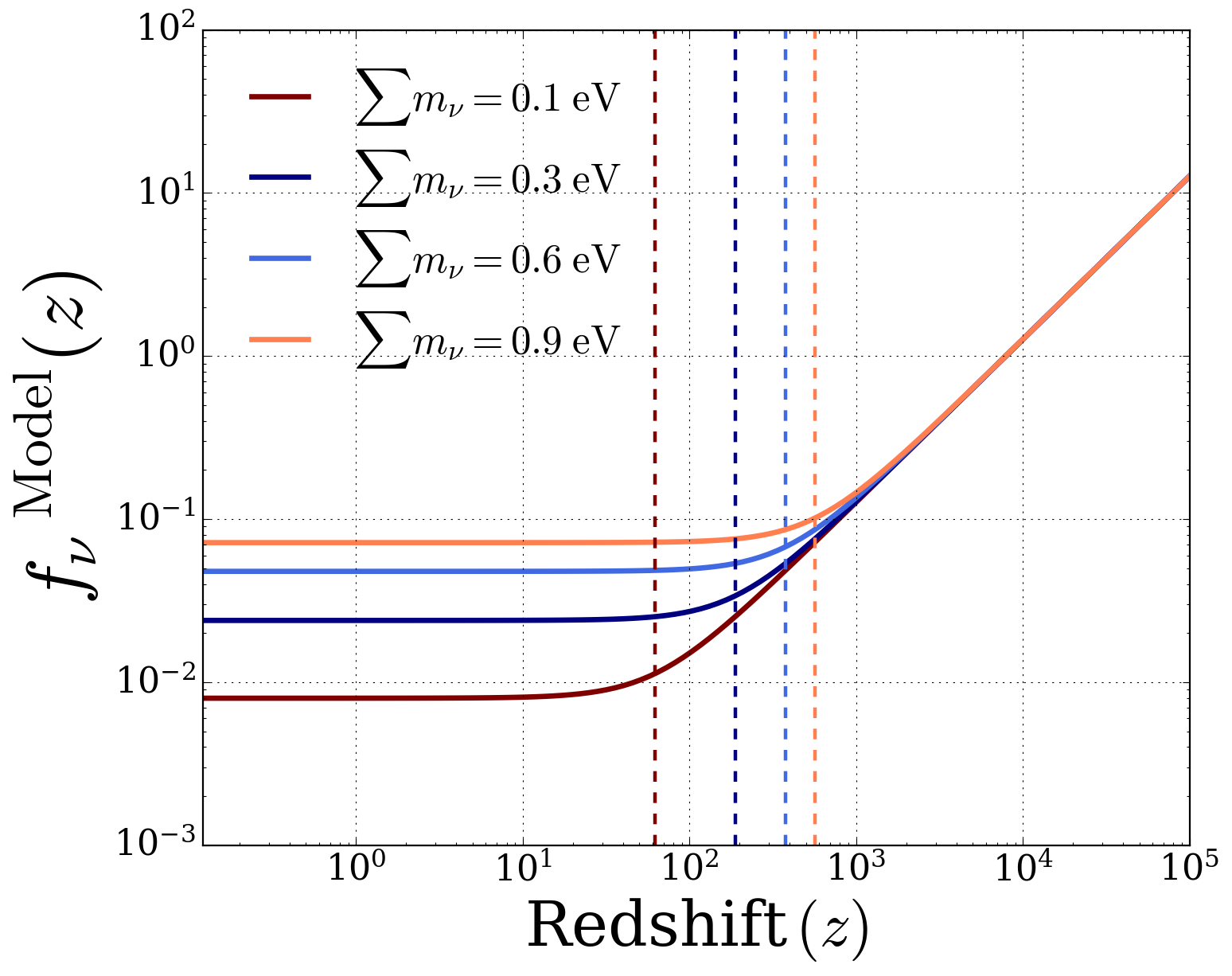}%
    \caption{\textit{Left:} The ratio of the neutrino mass fraction from our model to that obtained from \textsc{class} for several neutrino mass sums, shown as a function of redshift. Unlike the equation of state $w(z)$, our mass fraction disagrees notably with that from \textsc{class} at redshifts above $z_{\rm T}$. However, our approach is accurate for $z < z_{\rm T}$, which is where we wish to use these results. \textit{Right:} The neutrino mass fraction, $\fnu(z)$, vs. redshift $z$, in our model. $f_{\nu}$ stays constant after the neutrinos become non-relativistic. The vertical lines are the transition redshift corresponding to each neutrino mass.}%
    \label{fnu(z)}%
\end{figure}

Once the neutrinos become non-relativistic, $\fnu$ becomes a constant; the neutrinos are then counted as matter and so $\Omega_{\rm m} = \Omega_{\rm b}+\Omega_{\rm c}+\Omega_{\nu}$ where $\Omega_{\rm b}$, $\Omega_{\rm{c}}$, and $\Omega_{\nu}$ are the baryon, cold dark matter (CDM), and neutrino density fractions, respectively. It is also convenient to express the transition scale factor $a_{\rm T}$ in terms of the neutrino mass fraction at present, $\fnu$:
\begin{align}
    a_{\rm T} = \frac{1.70\times 10^{-5}}{\Omega_{{\rm m}}h^2}\fnu^{-1}.
\label{transitionaT}
\end{align}
Neutrino oscillation experiments indicate that at least two of the three neutrino flavors are non-relativistic today. Even if the third flavor is relativistic, equation (\ref{transitionaT}) is still a good approximation \cite{LESGOURGUES_2006}. In this paper, we assume three degenerate massive neutrino species. The approximate neutrino mass fraction vs. the exact results from \textsc{class} is displayed in Fig. \ref{fnu(z)}. The left panel shows the evolution of the mass fraction and the right panel is the ratio of the mass fraction to that of \textsc{class}. The plots show good agreement between the approximation in (\ref{fnu}) and exact solutions, especially for when the neutrinos are already non-relativistic.

\subsection{Boltzmann Equation \& Free-Streaming Scale}
\label{Boltzmann Equation}

Previous works \citep{shoji2010massive, 1995, Blas_2014} show how to obtain from the Boltzmann hierarchy the fluid equations for a massive neutrino and CDM cosmology.\footnote{A two-fluid approach (baryons, and CDM) was used in \cite{1994, bashinsky2001position,bashinsky2002dynamics} to yield a good approximation of the angular power spectrum for CMB anisotropies, and in \cite{2016MNRAS.457...24S, bashinsky2002dynamics} to obtain a fully analytic treatment of Baryon Acoustic Oscillations (BAO) in the density field. Neither of these works focus on neutrinos, so we merely note them here to point to additional work that has exploited the simplicity of a two-fluid approach.} Here we briefly summarize the idea behind the Boltzmann hierarchy. We start with the unperturbed phase space distribution function of particles, $f_0(q,\tau)$, which can be taken to be the Fermi-Dirac distribution (\ref{mean momentum}) for neutrinos and matter. Then we perturb the distribution function; the full distribution function ,$f(\vec{k},\hat{n},q,\tau)$, then becomes:
\begin{align}
    f(\vec{k},\hat{n},q,\tau) = f_0(q,\tau)\left(1+\Psi(\vec{k},\hat{n},q,\tau)\right)
\label{DF}
\end{align}
where $\vec{k}$ is the wave-vector of the perturbation modes, $\hat{n}\equiv \vec{q}/q$, the direction of the momentum of the particles, and $\Psi$ is the perturbation.

We next write the collisionless Boltzmann equation and obtain the evolution of $\Psi(\vec{k},\hat{n},q,\tau)$. Lastly, we expand $\Psi(\vec{k},\hat{n},q,\tau)$ in terms of Legendre polynomials with respect to $\mu \equiv \hat{k}\cdot \hat{n}$ and obtain an infinite series of coupled differential equations for each moment $\ell$ of the Legendre series. This set of equations is called the Boltzmann hierarchy. In practice, we need to truncate the hierarchy at some $\ell_{\rm max}$. For the neutrinos, as it turns out, once the ratio of $\epsilon/q$ is small, we can truncate the hierarchy at $\ell_{\rm max} = 1$ \cite{shoji2010massive}. Therefore, the evolution of the  density and velocity perturbations is sufficient for describing the system. 

We also assume that the vorticity of the cosmic fluid is negligible. This assumption is studied in \cite{2015}, which finds that when $k<1\,h\rm{Mpc}^{-1}$ the vorticity can be neglected. \cite{2014mathias} has shown that vorticity can be produced in two-loop Effective Field Theory of Large-Scale Structure, but this is relevant only on small scales. Therefore, it is safe to assume that the velocity field is curl-free for the scales of interest in the present work. 

Finally, we arrive at the linear fluid equations for cold dark matter (CDM) and baryons, where the sum of these is denoted with a subscript cb, and neutrinos, denoted with subscript $\nu$. In Fourier space, we have:
\begin{align}
    &\dot{\delta}_{\rm cb}(k,\tau)+\theta_{\rm cb}(k,\tau)-3\dot{\phi}(k,\tau)=0,\label{dcb}\\
    &\dot{\theta}_{\rm cb} (k,\tau) +\mathcal{H}(\tau)\theta_{\rm cb} (k,\tau)+\frac{3}{2}\mathcal{H}^{2}(\tau)\Omega_{\rm m}(\tau)\left( f_{\nu}\delta_{\nu}(k, \tau)+(1-f_{\nu})\delta_{\rm cb}(k, \tau) \right) = 0,\label{tcb}\\
    &\dot{\delta}_{\nu}(k, \tau)+\theta_{\nu}(k, \tau) -3\dot{\phi}(k,\tau)= 0,\label{dnu}\\
    &\dot{\theta}_{\nu}(k, \tau)+\mathcal{H}(\tau)\theta_{\nu}(k, \tau)+\frac{3}{2}\mathcal{H}^{2}(\tau) \Omega_{\rm m}(\tau)\Big(f_{\nu}\delta_{\nu}(k, \tau)\nonumber\\
   &\qquad\qquad\qquad\qquad\qquad\qquad +(1-f_{\nu})\delta_{\rm cb}(k, \tau) \Big) -k^2 {c_s}^2 \delta_{\nu}(k, \tau) 
= 0,
    \label{tnu}
\end{align}
where $\mathcal{H}$ is the Hubble parameter in terms of conformal time $d\tau \equiv dt/a(t)$, and dot denotes a derivative with respect to conformal time. $\delta$ and $\theta$ are the usual density perturbation and velocity divergence, in Fourier space. They are originally defined in position space via \cite{Bernardeau_2002}:
\begin{align}
    \delta(\vec{r},\tau)=\frac{\rho(\vec{r},\tau)}{\bar{\rho}(\tau)}-1
    \label{delta}
\end{align}
\begin{align}
    \theta(\vec{r},\tau)=\nabla\cdot\vec{v}(\vec{r},\tau)\label{theta}.
\end{align}
Finally, $\phi(k,\tau)$ is the gravitational potential in the Newtonian gauge \citep{1995, shoji2010massive} and is sourced by the matter density perturbations via the Poisson equation: 
\begin{align}
    k^2 \phi(k,\tau) = \frac{3}{2}\mathcal{H}^2(\tau)\Omega_{\rm m}(\tau)\delta_{\rm m}(k,\tau).
    \label{Poisson}
\end{align} 
For a geometrically flat, matter-dominated Universe, $\delta \propto a$ and so $\phi$ is a constant. We can make this assumption prior to dark energy domination, which begins at $z_{\rm{eq}} = 0.37$ and continues to lower redshift. Dark energy's impact on our calculation is explored further in \S\ref{Dark Energy Modifications}.

$c_s(z)$ is the neutrino sound speed and can be expressed as ${c_s}^2(z) = (5/3)\; w(z)$ \citep{shoji2010massive, Blas_2014}. The neutrino sound speed is understood best if it is converted to a length-scale, the \textit{free-streaming} scale, which is conceptually similar to the Jeans length:
\begin{align}
    {k_{\rm FS}}^2(\tau) \equiv \frac{3}{2}\Omega_{\rm m}(\tau) \frac{\mathcal{H}^2(\tau)}{{c_ {s}}^2(\tau)}=\frac{27}{10}\frac{\Omega_{\rm m}H_0^2}{c^2}\frac{1}{a(\tau)} \left(1+\left(\frac{a(\tau)}{a_{\rm T}}\right)^2 \right),
\label{kfree}
\end{align}
We note that this equation is general, even if one did not have a matter-dominated cosmology (for instance, if we added dark energy); this occurs because the time dependence in $\Omega_{\rm m}(\tau)$ cancels with that in $\mathcal{H}(\tau)$. We also note that after the first equality, $\Omega_{\rm m}$ is a constant, the present time value of $\Omega_{\rm m}(\tau)$.

It is clear that if $a>\aT$, the second term in the parentheses in (\ref{kfree}) is much larger than 1 and therefore we can take ${k_{\rm FS}}^2(\tau) \propto a$. As a result, we can define the free-streaming scale at present as:
\begin{align}
    k_0 = \left(\frac{27}{10}\frac{H_0^2\Omega_{{\rm m}}}{c^2{a_{\rm T}}^2}\right)^{\frac{1}{2}} = 0.0113\, \left(\frac{\Omega_{\rm m}}{0.3}\right)^{\frac{1}{2}} \left(\frac{\sum m_{\nu}}{0.06 \; \rm eV}\right)\, [h\rm{\,Mpc^{-1}}]
\label{kfree0}
\end{align}
and the maximum free-streaming scale which is formally defined as the \textit{non-relativistic} scale $k_{\rm NR}$ as the free-streaming scale at the time of transition:
\begin{align}
   {k_{\rm{NR}}} = \sqrt{2 {k_0}^2 a_{\rm T}} =0.133\; k_0 \left(\frac{a_{\rm T}}{1/113.5} \right)^{\frac{1}{2}}= 0.00261\left( \frac{\Omega_{\rm m}}{0.3}\right)^{\frac{1}{2}} \left(\frac{\sum m_{\nu}}{0.06 \; \rm eV} \right)^{\frac{1}{2}}\;[h\rm{\,Mpc^{-1}}],
   \label{kNR}
\end{align}
where we have also assumed three degenerate neutrino masses, \textit{i.e.} $\sum m_{\nu} = 3m_{\nu}$. The free-streaming scale is a characteristic length scale below which $( \textit{i.e.}\; k \gg k_{\rm FS})$ the neutrinos do not cluster due to their high thermal velocities, and therefore do not contribute to the growth of matter perturbations. On the other hand, on larger scales $(\textit{i.e.}\;k \ll k_{\rm FS})$, the neutrinos behave like CDM and so there is no observable effect on the matter perturbations \cite{Levi16}. Thus we expect that most of the neutrino effects of interest occur on scales between $k_{\rm NR}$ and $k_0$. The shape of the power spectrum on these scales will uniquely constrain the neutrino mass fraction $\fnu$. 
 
\section{Two-Fluid Equations}
\label{Two-fluid equations}

We combine the system of equations given by (\ref{dcb})-(\ref{tnu}) to arrive at the two-fluid system of equations for neutrinos and CDM+baryons. Assuming that the gravitational potential is almost constant, we find:
\begin{align}
    &{\delta}_{\rm{cb}}^{''} (k, s)+\left(\frac{{\mathcal{H}}^{'}}{\mathcal{H}}(s)+1\right){\delta}_{\rm{cb}}^{'}(k, s) = \frac{3}{2}\Omega_{\rm{m}}(s)\left( f_{\nu}\delta_{\nu}(k,s)+(1-f_{\nu}){\delta}_{\rm{cb}}(k,s) \right),\label{dcb''}\\
    &\dnu^{''}(k,s)+\left(\frac{{\mathcal{H}}^{'}}{\mathcal{H}}(s)+1\right)\dnu^{'}(k,s) \nonumber\\
    &\qquad\qquad \qquad= \frac{3}{2}\Omega_{\rm{m}}(s)\left(f_{\nu}  \delta_{\nu}(k,s)+(1-f_{\nu}) {\delta}_{\rm{cb}}(k,s)-\left(\frac{k}{k_{\rm{FS}}(s)}\right)^{2} \dnu(k,s) \right),\label{dnu''}
\end{align}
where the derivatives are with respect to  $s \equiv \ln{a}$. We emphasize that we have not considered the neutrino mass fraction to be time-dependent. We are explicitly interested in redshifts later than the transition and at those times, the neutrinos count as part of the matter. The total matter density perturbation is therefore:
\begin{align}
    \dm(k,s) = \fnu \dnu(k,s) + \left(1-\fnu\right) \dcb(k,s).
\label{matterDP}
\end{align}

\section{Iterative Method}
\label{Iterative Method}
\subsection{Setup}
\label{Set-up}
We now describe how we solve the system of equations (\ref{dcb''}) and (\ref{dnu''}). For simplicity, we assume that the Universe is matter-dominated, so that we can take $\Omega_{\rm m}(\tau) = 1$; we also assume that the neutrinos do not change the rate at which the Universe expands after the non-relativistic transition; this leads to $\mathcal{H}^{'}/\mathcal{H}+1 = 1/2$. 

Our approach will be  to expand the system of equations (\ref{dcb''}) and (\ref{dnu''}) around $\fnu=0$ and solve perturbatively order by order in $\fnu$, working up to $\mathcal{O}(\fnu)$. We have, suppressing the spatial and time dependence:
\begin{align}
    \dcb &= \dcb^{(0)}+\dcb^{(1)}+\cdots,\\
\nonumber\\ 
    \dnu &= \dnu^{(0)}+\dnu^{(1)}+\cdots,
\label{dnuexpand}
\end{align}
where $\delta_i^{(n)}$ is a perturbation to the $i^{th}$ species at order $n$ in $\fnu$. 

Since we want to determine the impact specifically of the neutrinos' clustering on the matter, we focus on times $a>a_{\rm{T}}$.\footnote{Prior to this, on scales $k > k_{\rm f s}$ the neutrinos form a smooth background and simply suppress the growth of structure by contributing to the Hubble parameter but not the density perturbations, exactly like the M\'esz\'aros effect \citep{Meszaros:1974}.} The zeroth-order equations are then:
\begin{align}
    &{\delta}_{\rm cb}^{{(0)}^{''}} (q, s)+\frac{1}{2}{\delta}_{\rm cb}^{{(0)}^{'}}(q, s) - \frac{3}{2}\dcb^{(0)}(q, s) = 0\label{dcb0order},\\
\nonumber\\
    &\dnu^{{(0)}^{''}}(q, s)+\frac{1}{2}\dnu^{{(0)}^{'}}(q, s)+{3} q^2 e^{s_{\rm T}-s}\dnu^{(0)}(q, s) - \frac{3}{2}\dcb^{(0)}(q, s)=0,
\label{dnu0order}
\end{align}
where we have introduced $q \equiv k/k_{\rm  NR}$ (as in \ref{kNR}) and also defined  $s_{\rm T} \equiv \ln{a_{\rm T}}$.

We highlight that our zeroth order system above is coupled---the neutrino perturbation equation receives an $\mathcal{O}(\fnu^0)$ contribution from $\delta_{\rm cb}$, though the equation for $\delta_{\rm cb}$ does not receive a contribution from $\delta_{\nu}$ (this latter enters only at $\mathcal{O}(\fnu)$). We note that the zeroth-order CDM+baryon equation (\ref{dcb0order}) is a homogeneous differential equation while the neutrino zeroth-order (\ref{dnu0order}) is an inhomogeneous equation with the CDM+baryons as the external forcing term.


Now, the first-order equations, \textit{i.e.} linear in $\fnu$, are:
\begin{align}
    &{{\delta}_{\rm cb}^{(1)}}^{''}(q, s)+\frac{1}{2}{{\delta}_{\rm cb}^{(1)}}^{'}(q, s)-\frac{3}{2}\dcb^{(1)}(q, s)  = \frac{3}{2}\fnu \left( \dnu^{(0)}(q, s) -\dcb^{(0)}(q, s)\right) \equiv F^{(0)}(q, s),\label{dcb1order}\\
\nonumber\\
    &{\dnu^{(1)}}^{''}(q, s)+\frac{1}{2}{\dnu^{(1)}}^{'}(q, s)+{3}q^2 e^{s_{\rm T}-s}\dnu^{(1)}(q, s) =  \frac{3}{2}\dcb^{(1)}(q, s) - \fnu \dcb^{(0)}(q, s).
\label{dnu1order}
\end{align}
We highlight that the system above is both coupled ($\delta^{(1)}_{\rm cb}$ enters the equation for $\delta_{\nu}^{(1)}$) and also externally forced, by the zeroth-order perturbations $\delta_{\nu}^{(0)}$ and $\delta_{\rm cb}^{(0)}$. The forcing term defined as $F^{(0)}(q, s)$ in (\ref{dcb1order}) is sensitive to the difference between the neutrino and matter density perturbations. On large scales, this difference would vanish at late times, and on small scales, the neutrino density perturbations are suppressed and therefore, they will not contribute to the forcing term. The homogeneous part of (\ref{dcb1order}) is exactly the same as that of (\ref{dcb0order}), so the homogeneous solutions will be the same. 

The general form of the solution for an inhomogeneous differential equation with a forcing term is given by variation of parameters as:
\begin{align}
    \delta_{\rm cb}^{(1)}(q, s) = G(s)\left(c_G(q) - \int_{s_{\rm T}}^{s}\frac{D(s^{'})F(q, s^{'})}{W(q, s^{'})}ds^{'}\right)+D(s)\left(c_D(q) - \int_{s_{\rm T}}^{s}\frac{G(s^{'})F(q, s^{'})}{W(q, s^{'})}ds^{'}\right)
    \label{GeneralForm}
\end{align}
where $c_G(q)$ and $c_D(q)$ are the initial conditions, $G(s)$ and $D(s)$ are the growing and decaying solutions to the homogeneous equation, and $W(q,s) = G(s)D^{'}(s) - D(s)G^{'}(s)$ is the Wronskian.  We can observe that if the initial conditions are zero (which is the case for the first-order CDM+baryons equation, \ref{dcb1order}), the solution will come only from the integral. 

In summary, inspecting (\ref{dcb0order})-(\ref{dnu1order}) shows that the zeroth-order matter perturbations will source the zeroth-order neutrino perturbations. The neutrinos then change the way the matter evolves according to (\ref{dcb1order}). Since we are only interested in the lowest-order solutions in $\fnu$, we only have to solve the zeroth-order equations for each species and then also the first-order one for the CDM+baryons, (\ref{dcb1order}). We do not need to solve the first-order equation for the neutrino perturbation.


\subsection{Solutions}
\label{solutions}
The solution to equation (\ref{dcb0order}) for $\dcb^{(0)}$ is:
\begin{align}
    {\delta}_{\rm cb}^{(0)}(q,s) = c_G(q)e^s+c_D(q) e^{-3s/2},
\label{0orderdcb}
\end{align}
where $c_G$ and $c_D$ are the initial conditions to the growing and decaying modes, $G(s)=e^s$ and $D(s) = e^{-3s/2}$, respectively. Ignoring the decaying mode, we may write  $\dcb^{(0)}(q,s) = \mathcal{N}\dcb^{\fnu=0}(q,0) e^s$ where $\mathcal{N}$ is a normalization factor and $\dcb^{\fnu=0}(q,0)$ is the present-time CDM density perturbation in the absence of neutrinos. We define $\dcb(q,0) \equiv \mathcal{N}\dcb^{\fnu=0}(q,0)$. The initial conditions for the neutrino equation (\ref{dnu0order}) are taken to be $\delta_{\nu}(q, s_{\rm T}) = \delta_{\nu}^0(q)$ and $\delta_{\nu}^{'}(q, s_{\rm T}) = \theta_{\nu}^0(q)$. These initial conditions are taken to be  smoothed Heaviside functions that are equal to the matter density and velocity perturbations at the transition redshift $\delta_{\rm cb}(q, s_{\rm T})$ and $\theta_{\rm cb}(q, s_{\rm T})$ for $k<k_{\rm NR}$. We set the initial conditions to zero for $k>k_{\rm NR}$ since neutrinos are free-streaming below the non-relativistic scale. 


We find the solution to (\ref{dnu0order}), for $\dnu^{(0)}(q,s)$, as:
\begin{empheq}[box=\fbox]{align}
    \dnu^{(0)}(q,s) &=  \delta_{\rm cb}(q, 0) Q^2 \Bigg[ \cos \left( Q e^{-s/2} \right) \left\{ {\rm Ci}\left(Q e^{-s/2} \right)-{\rm Ci}\left(Q e^{-s_{\rm T}/2} \right)\right\}\nonumber\\
    &\qquad\qquad\qquad+ \sin \left(Q e^{-s/2}\right) \left\{ {\rm Si}\left(Q e^{-s/2} \right)-{\rm Si}\left(Q e^{-s_{\rm T}/2} \right)\right\}\Bigg]\nonumber\\
    &+\left(\delta_{\nu}^{0}(Q)-\delta_{\rm cb}(q, 0) e^{s_{\rm T}}\right) \cos \left(Q\left(e^{-s/2}-e^{-s_{\rm T}/2}\right)\right)\nonumber\\
    & + Q^{-1} e^{s_{\rm T}/2} \left[ \delta_{\rm cb}(q, 0) \left\{2 e^{s_{\rm T}}- Q^2\right\}-2 \theta_{\nu}^{0}(Q)\right] \sin \left(Q \left(e^{-s/2}-e^{-s_{\rm T}/2}\right)\right)\nonumber\\
    &+\delta_{\rm cb}(q, 0) e^s,
    \label{dnu_complete}\\
    &\qquad \qquad \qquad Q \equiv 2 \sqrt{3 e^{s_{\rm T}}}\, q, \nonumber
\end{empheq}
with ${\rm Ci}$ and ${\rm Si}$ respectively the Cosine and Sine integrals. 
As we see from (\ref{dnu_complete}), the terms that are proportional to the neutrino initial conditions only oscillate around $\delta_{\nu}^0(Q)$ and $\theta_{\nu}^0(Q)$ and therefore, they do not grow. Hence, at late time $z<10$ we may neglect these initial conditions. We also note that on small scales where $q\gg1$, the arguments of $\rm Si$ and $\rm Ci$ functions are very large, which means we may asymptotically expand them in terms of $Q$.

Now with the zeroth-order solution of the neutrinos, we may solve the CDM+baryons at first order, \textit{i.e.} solve (\ref{dcb1order}). The homogeneous solution to (\ref{dcb1order}) is the exact growing and decaying modes as the zeroth-order CDM+baryons solutions (\ref{0orderdcb}). However, since the initial conditions on the density and velocity perturbations of (\ref{dcb1order}) are both zero, the solutions are only produced from the integral in (\ref{GeneralForm}). We find:
\begin{align}
    \delta_{\rm cb}^{(1)}(k, s) = {\delta
    }_{\rm I}(k,s) + \delta_{\rm II}(k,s)+\delta_{\rm int}(k,s)
\end{align}

where we have defined ${\delta}_{\rm I}(k,s)$, $\delta_{\rm II}(k,s)$ and $\delta_{\rm int}(k,s)$ as:
\begin{align}
    {\delta}_{\rm I}\left(k,s\right) &= 
    \frac{6 f_{\nu}}{5 Q^3}e^{\frac{s}{2}}\nonumber\\&\times\Bigg[  Q e^{\frac{s}{2}} \left( \delta_{\nu}^0(Q)+\delta_{\rm cb}\left(q, 0\right) Q^2 \left(\frac{s_{\rm T}-s}{2}+1\right)-3 \delta_{\rm cb}\left(q, 0\right) e^{s_{\rm T}}+2 \theta_{\nu}^0(Q)\right)\nonumber\\&\quad- \Big(\delta_{\nu}^0(Q) Q^2+e^{\frac{s+s_{\rm T}}{2}} \left(-\delta_{\rm cb}\left(q, 0\right) Q^2+2 \delta_{\rm cb}\left(q, 0\right) e^{s_{\rm T}}-2 \theta_{\nu}^0(Q)\right)-\delta_{\rm cb}\left(q, 0\right) Q^2 e^{s_{\rm T}}\Big)\nonumber\\ &\qquad\times \sin \left(Q \left(e^{-\frac{s}{2}}-e^{-\frac{s_{\rm T}}{2}}\right)\right)\nonumber\\ &\quad- \Bigg(\delta_{\nu}^0(Q) e^{s/2}+e^{s_{\rm T}/2} \left(2 \theta_{\nu}^0(Q)-\delta_{\rm cb}\left(q, 0\right) \left(-Q^2+e^{\frac{s+s_{\rm T}}{2}}+2 e^{s_{\rm T}}\right)\right)\Bigg)\nonumber\\&\qquad \times Q \cos \left(Q \left(e^{-\frac{s}{2}}-e^{-\frac{s_{\rm T}}{2}}\right)\right) \Bigg],
\end{align}
\begin{align}
    \delta_{\rm II}\left(k,s\right) &= \frac{ 6 f_{\nu}}{5}     \delta_{\rm cb}\left(q, 0\right)  e^{s/2} \nonumber\\&\times \Bigg[{\rm Ci}\left(Qe^{-\frac{s_{\rm T}}{2}} \right) \left(Q \sin \left(Q e^{-\frac{s}{2}}\right)+e^{s/2} \cos \left(Q e^{-\frac{s}{2}}\right)\right)\nonumber\\&\quad-{\rm Ci}\left(Q e^{-\frac{s}{2}} \right) \left(Q \sin \left(Q e^{-\frac{s}{2}}\right)+e^{s/2} \cos \left(Q e^{-\frac{s}{2}}\right)\right)\nonumber\\&\quad+\left({\rm Si}\left(Q e^{-\frac{s}{2}} \right)-{\rm Si}\left(Q e^{-\frac{s_{\rm T}}{2}} \right)\right) \left(Q \cos \left(Q e^{-\frac{s}{2}}\right)-e^{s/2} \sin \left(Q e^{-\frac{s}{2}}\right)\right)\Bigg],
\end{align}
\begin{align}
    \delta_{\rm int}(k,s) &= \frac{3f_{\nu}}{5 Q}e^{-5s/2}\nonumber\\
    &\times\int_{s_{\rm T}}^{0}e^{3t/2} \Bigg[Q^3 \Bigg(\cos \left(Q e^{-t/2} \right) \left({\rm Ci}\left(Q e^{-s_{\rm T}/2} \right)-{\rm Ci}\left(Q e^{-t/2} \right)\right)\nonumber\\
&\qquad\qquad\qquad\quad+\sin \left(Q e^{-t/2} \right) \left({\rm Si}\left(Q e^{-s_{\rm T}/2} \right)-{\rm Si}\left(Q e^{-t/2} \right)\right)\Bigg)\nonumber\\
&\qquad\qquad+Q \left(e^{s_{\rm T}}-\delta_{\nu}^0(Q)\right) \cos \left(Q \left(e^{-t/2}-e^{-s_{\rm T}/2}\right)\right)\nonumber\\
    &\qquad\qquad+e^{s_{\rm T}/2} \left(-Q^2+2 e^{s_{\rm T}}-2 \theta_{\nu}^0(Q)\right) \sin \left(Q \left(e^{-s_{\rm T}/2}-e^{-t/2}\right)\right)\Bigg]dt.
\end{align}

\subsection{Closed-Form Power Spectrum Including Neutrino Mass}
Next, we combine the solutions from the matter and neutrinos according to (\ref{matterDP}) to find the matter density perturbation at first order in the neutrino mass fraction, $f_{\nu}$. By definition, the total matter power spectrum is:
\begin{align}
    P(k,\tau) = \left<\delta_{\rm m}(\vec{k}_1,\tau)\delta_{\rm m}(\vec{k}_2,\tau) \right>(2\pi)^3 \delta_{\rm D}^{[3]}(\vec{k}_1+\vec{k_2}),
\label{PS}
\end{align}
where $\left<\right>$ is the ensemble average. The matter power spectrum then  becomes (switching our time variable from $\tau$ to $s$):
\begin{align}
    P(k, s) = P_{f_{\nu}=0}(k, s)\left(1 + \mathcal{P}_{\nu}(k, s) + \mathcal{P}_{\rm Os.I}(k, s) + \mathcal{P}_{\rm Os.II}(k,s) + \mathcal{P}_{\rm int}(k,s)\right).
    \label{closed_form}
\end{align}
$P_{f_{\nu}=0}(k, s)$ is the total matter power spectrum in the absence of neutrinos, $\mathcal{P}_{\nu}(k, s)$ is the contribution due to linear neutrino density, $\mathcal{P}_{\rm Os.I}(k, s)$ and $\mathcal{P}_{\rm Os.II}(k,s)$ correspond to the oscillatory behavior of the neutrinos on small scales and lastly, $\mathcal{P}_{\rm int}(k,s))$ is the one-dimensional integral that needs to be evaluated numerically. The contributions are:
\begin{align}
  \mathcal{P}_{\nu}(k, s)  = 2 f_{\nu} e^{-s} &\Bigg[Q^2 \Bigg(\cos \left(Q e^{-s/2}\right) \left({\rm Ci}\left(e^{-s/2} Q\right)-{\rm Ci}\left(e^{-s_{\rm T}/2} Q\right)\right)\nonumber\\&\qquad+\sin \left(Q e^{-s/2}\right) \left({\rm Si}\left(e^{-s/2} Q\right)-{\rm Si}\left(e^{-s_{\rm T}/2} Q\right)\right)\Bigg)\nonumber\\&+\left(\delta_{\nu}^0(Q)-e^{s_{\rm T}}\right) \cos \left(Q \left(e^{-s/2}-e^{-s_{\rm T}/2}\right)\right)
  \nonumber\\&+\frac{1}{Q}e^{s_{\rm T}/2} \left(-Q^2+2 e^{s_{\rm T}}-2 \theta_{\nu}^0(Q)\right) \sin \left(Q \left(e^{-s/2}-e^{-s_{\rm T}/2}\right)\right)\Bigg],
\end{align}
\begin{align}
    \mathcal{P}_{\rm Os.I}(k, s) &= \frac{12 f_{\nu}}{5 Q^3} e^{-s/2} \\&\times \Bigg[Q e^{s/2} \left(\delta_{\nu}^0(Q)+Q^2 \left(1-\frac{s-s_{\rm T}}{2}\right)-3 e^{s_{\rm T}}+2 \theta_{\nu}^0(Q)\right)\nonumber\\&\quad-\sin \left(Q \left(e^{-s/2}-e^{-s_{\rm T}/2}\right)\right) \left(\delta_{\nu}^0(Q) Q^2+e^{\frac{s+s_{\rm T}}{2}} \left(-Q^2+2 e^{s_{\rm T}}-2 \theta_{\nu}^0(Q))+Q^2 (-e^{s_{\rm T}}\right)\right)\nonumber\\&\quad- Q \cos \left(Q \left(e^{-s/2}-e^{-s_{\rm T}/2}\right)\right) \left(\delta_{\nu}^0(Q) e^{s/2}+e^{s_{\rm T}/2} \left(Q^2-e^{\frac{s+s_{\rm T}}{2}}-2 e^{s_{\rm T}}+2 \theta_{\nu}^0(Q)\right)\right)\Bigg]\nonumber,
\end{align}
\begin{align}
    \mathcal{P}_{\rm Os.II}(k, s) &= \frac{12 f_{
\nu}}{5} e^{-s/2}\\&\times \Bigg[{\rm Ci}\left(e^{-s_{\rm T}/{2}} Q\right) \left(Q \sin \left(Q e^{-s/2}\right)+e^{s/2} \cos \left(Q e^{-s/2}\right)\right)\nonumber\\&\quad-{\rm Ci}\left(e^{-s/2} Q\right) \left(Q \sin \left(Q e^{-s/2}\right)+e^{s/2} \cos \left(Q e^{-s/2}\right)\right)\nonumber\\&\quad+\left({\rm Si}\left(e^{-s/2} Q\right)-{\rm Si}\left(e^{-s_{\rm T}{2}} Q\right)\right) \left(Q \cos \left(Q e^{-s/2}\right)-e^{s/2} \sin \left(Q e^{-s/2}\right)\right)\Bigg],\nonumber
\end{align}
and lastly:
\begin{align}
    \mathcal{P}_{\rm int}(k,s) &= \frac{6f_{\nu}}{5 Q}e^{-5s/2}\\
    &\times\int_{s_{\rm T}}^{0}e^{3t/2} \Bigg[Q^3 \Bigg(\cos \left(e^{-t/2} Q\right) \left({\rm Ci}\left(e^{-s_{\rm T}/2} Q\right)-{\rm Ci}\left(e^{-t/2} Q\right)\right)\nonumber\\
&\qquad\qquad\qquad\quad+\sin \left(e^{-t/2} Q\right) \left({\rm Si}\left(e^{-s_{\rm T}/2} Q\right)-{\rm Si}\left(e^{-t/2} Q\right)\right)\Bigg)\nonumber\\
&\qquad\qquad+Q \left(e^{s_{\rm T}}-\delta_{\nu}^0(Q)\right) \cos \left(Q \left(e^{-t/2}-e^{-s_{\rm T}/2}\right)\right)\nonumber\\
    &\qquad\qquad+e^{s_{\rm T}/2} \left(-Q^2+2 e^{s_{\rm T}}-2 \theta_{\nu}^0(Q)\right) \sin \left(Q \left(e^{-s_{\rm T}/2}-e^{-t/2}\right)\right)\Bigg]dt.\nonumber
\end{align}

$\mathcal{P}_{\rm Os. I}(k, s)$ and $\mathcal{P}_{\rm Os. II}(k, s)$ are highly oscillatory on scales smaller than $k_0$. However, since they are out of phase with each other, their oscillations cancel. $\mathcal{P}_{\rm int}(k,s)$ is also small in amplitude compared to the other terms and may be neglected if desired; however, we retain it throughout this paper.

\begin{figure}%
    \centering
    \subfloat{\includegraphics[width=15cm]{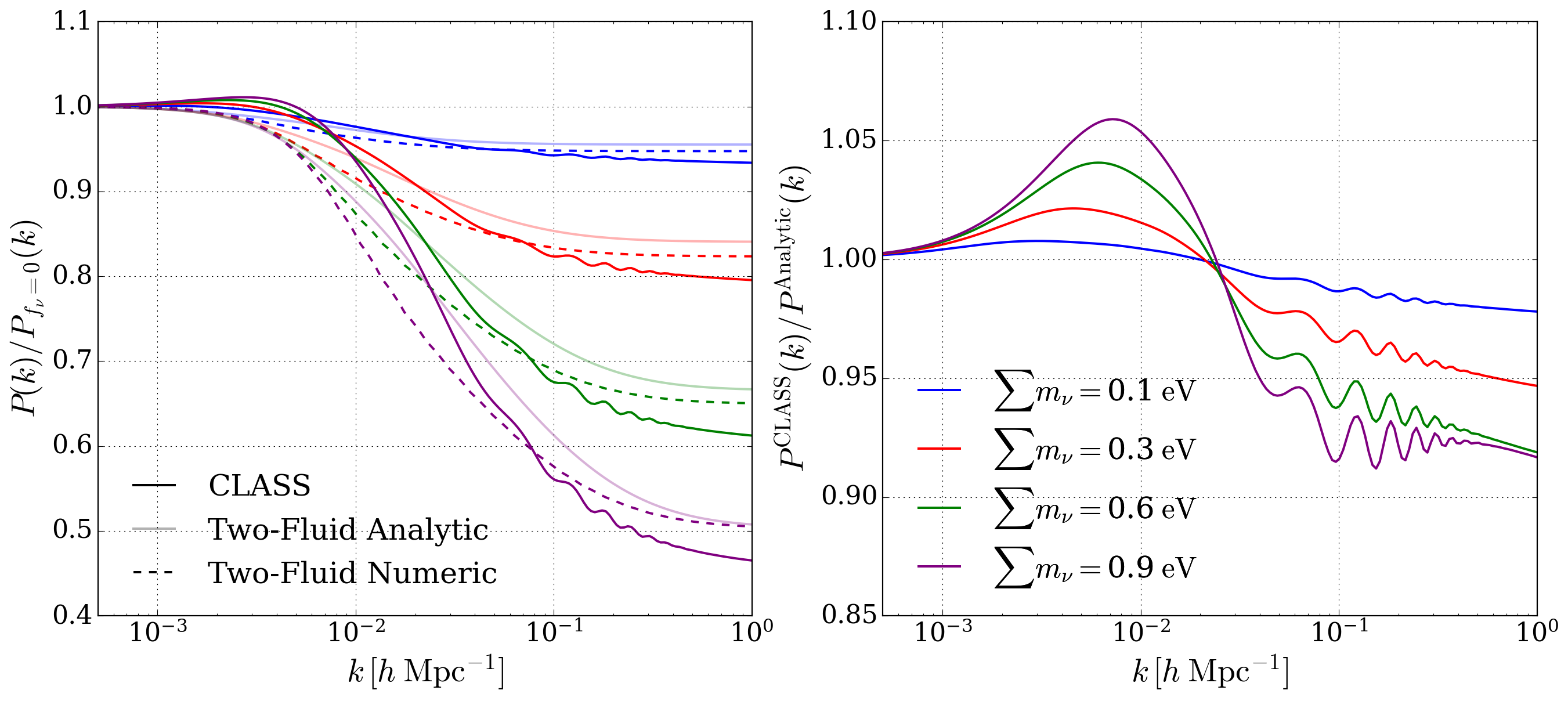} }%
    \caption{\textit{Left:} The ratio of the power spectrum with massive neutrinos, $P(k)$, to the power spectrum without massive neutrinos, $P_{\fnu=0}(k)$. The higher opacity solid curves show \textsc{class} results, while the lower opacity solid curves represent the analytic  model we obtained in (\ref{closed_form}). The dashed curves are from full numerical solution of our two-fluid system (\ref{dcb})-(\ref{tnu}). \textit{Right:} The ratio of the power spectrum from \textsc{class} to the analytic  power spectrum of this work (\ref{closed_form}). Our work shows very good agreement with \textsc{class}. The small differences between the two models are due, first, to neglecting the effects of dark energy on the expansion rate and on the gravitational potential (\ref{Poisson}), and second, to the approximations we made regarding the free-streaming scale.}%
    \label{fig:All three together}%
\end{figure}

(\ref{closed_form}) is the most important result of this work. We present a closed-form formula to calculate the matter power spectrum in presence of massive neutrinos from a simple and physically-motivated model for the equation of state (\ref{EOSparameter}). Also, our analytic  approach enables one to come up with an approximation to the cross-power spectrum between neutrinos and CDM+baryons, $P_{\rm cb\nu}$, which is required in some applications, such as calculating the non-linear corrections to the matter power spectrum at one-loop \cite{Wong_2008}. This approach may well accelerate the computation time in Fisher forecasting. 
The full-shape matter power spectrum suppression defined as $\Delta(k, s) = P(k,s)/P_{f_{\nu}=0}(k,s)$ is shown in Fig. \ref{fig:All three together} and demonstrates that the power spectrum from our analytic calculation (lighter-colored curves) compares quite favorably with the results from \textsc{class} (darker-colored curves). We also show the full two-fluid numerical solution, in dashed.

Our approximation achieves better than $5\%$ agreement for $k<k_0$. For a more realistic neutrino mass of $\sum m_{\nu} < 0.26\; {\rm eV}$, the agreement between our analytic  model and the exact solution from \textsc{class} improves to approximately $2\%$ on these scales.
\section{Limiting Form on Small Scales}
\label{limiting form}
\subsection{Asymptotic Solutions}
The closed-form formula obtained in the previous section (\ref{closed_form}) is valid on all scales. However, it is also necessary to understand the behavior of the matter power spectrum on very large scales, $k > k_0$. To determine this behavior, we repeat the procedure described previously. Instead of using the full solution for the zeroth-order neutrinos to (\ref{dnu0order}), we asymptotically expand it for $Q \gg 1$. Performing the expansion in $Q$, On large scales we find:
\begin{align}
    \dnu^{(0)}(Q \ll1,s) & =\dcb(q,0)e^s.\label{dnusmallq}
\end{align}
and on small scales:
\begin{align}
    \dnu^{\left(0\right)}\left(Q\gg1,s\right) &=
    \delta_{\nu}^0(Q) \cos \left(2 Q \left(e^{-\frac{s}{2}}-e^{-\frac{s_{\rm T}}{2}}\right)\right)-\frac{3 \delta_{\rm cb}\left(q, 0\right) e^{2 s_{\rm T}} \cos \left(2 Q \left(e^{-\frac{s}{2}}-e^{-\frac{s_{\rm T}}{2}}\right)\right)}{2 Q^2}\nonumber\\&+\frac{3 \delta_{\rm cb}\left(q, 0\right) e^{2 s}}{2 Q^2}-\frac{e^{s_{\rm T}/2} \theta_{\nu}^0(Q) \sin \left(2 Q \left(e^{-\frac{s}{2}}-e^{-\frac{s_{\rm T}}{2}}\right)\right)}{Q}
\label{dnulargeq}
\end{align}
It is important to note that we made no assumptions about the suppression of density perturbations on small scales---rather, this emerges naturally from our two-fluid scheme. 
(\ref{dnulargeq}) describes the neutrino behavior on small scales  and as we see, the only growing, non-oscillatory contribution is third one. Now if we ignore the oscillations and initial conditions simply because they are not growing, we may obtain a simpler expression.
Since $q=k/k_{\rm NR}$, we may re-write (\ref{dnulargeq}) as a function of $k_{\rm NR}$ to find:
\begin{align}
    \dnu^{(0)}(k\gg k_{\rm NR},s\gg s_{\rm T}) =\frac{1}{8}\delta_{\rm cb}(k, s) e^{s-s_{\rm T}}\left(\frac{k_{\rm NR}}{k}\right)^2.
\end{align}

Our linear model (in both $\fnu$ and the perturbations $\delta$) indicates that the neutrino perturbations are suppressed by a factor of $(k_{\rm NR}/k)^2$ on scales smaller (higher $k$) than $k_{\rm NR}$. As we go to higher wave-numbers, this suppression is more drastic. 
On non-linear scales $k_{\rm{NL}}$ the suppression is of order $(k_{\rm{NR}}/k_{\rm NL})^2\ll 1$ which is significant enough so that we may take the neutrino perturbations to be not-affected by the small-scale dynamics.  This is the assumption on which the first non-linear treatments of structure formation were built \citep{Saito_2008,Wong_2008}. Later work showed that this assumption violates momentum conservation on small scales \cite{Blas_2014}. In this work, we restrict ourselves to the linear effects of the neutrinos.

We now obtain $\dcb^{(1)}$ in the large-$k$ limit. Ignoring the decaying modes, the dominant contribution to the  asymptotic solution of (\ref{dcb1order}) is:
\begin{align}
    \dcb^{(1)}(k\gg k_{\rm NR},s) &=\frac{6}{25}\delta_{\rm cb}(k,0)e^{s}f_{\nu}\Bigg[1+\frac{25  e^{s-s_{\rm T}}}{112 }\left(\frac{k_{\rm NR}}{k}\right)^2+\frac{5}{2}   (s_{\rm T}-s)  \Bigg].\label{dcblargeq2}
\end{align}


The total matter density perturbation on small scales $(q\rightarrow\infty)$, is obtained by using (\ref{0orderdcb}), (\ref{dnulargeq}) and (\ref{dcblargeq2}):
\begin{align}
    \dm(k,s) =\dcb(k,0) e^s&\left[1-\frac{19}{25}\fnu+ \left(\frac{3}{5}\left( s_{\rm T}- s\right)+\frac{5}{28}e^{s-s_{\rm T}}\left(\frac{k_{\rm NR}}{k}\right)^2\right)\fnu\right].
\label{dm}
\end{align}

\subsection{Growth-Rate \& Power Spectrum on Small Scales}

Let us see whether this simple model can re-create the approximations that was previously obtained in the literature. It has been shown that the linear growth rate of structures in presence of neutrinos will be different from a universe without massive neutrinos. Therefore, the growth-rate can be used to probe the neutrino mass with future galaxy surveys \cite{hernandez2017neutrino}.

From the above, we may compute the linear matter growth rate (post-transition) from the above and compare to the fitting formulae of \cite{1999} and \cite{1998-smallscale}. Neglecting terms that are of order $\left(k_{\rm NR}/k\right)^2$ and Taylor-expanding when $\fnu$ is small we find from (\ref{dm}) that
\begin{align}
    D(a) \approx D_{\fnu=0}(a) \left[1-\frac{3}{5}\fnu\ln{\left(\frac{D_{\fnu=0}(a)}{D_i}\right)}\right] \propto D_{\fnu=0}^{1-\mu}(a)
\label{dmlargeq}
\end{align}
where $\mu \equiv 3/5f_{\nu}$.   (\ref{dmlargeq}) is therefore consistent with previous work \cite{Hu_1998}.
To obtain the asymptotic power spectrum on small scales, we may use (\ref{dm}) along with the definition of the power spectrum, (\ref{PS}). We have already seen that the neutrinos do not change the matter density perturbations on large scales; consequently this part of the power spectrum is completely insensitive to the neutrino mass. On small scales, however, we the neutrinos change $\dm(q,s)$ according to (\ref{dm}). Therefore we calculate the power spectrum at $s=0$ (the present) as:
\begin{align}
    P(q,0) = \mathcal{N}^2 P_{\fnu=0}(q,0)&\left[1-\frac{38}{25}\fnu+ \left(\frac{6}{5} s_{\rm T}+\frac{5}{14}e^{-s_{\rm T}}
    \left(\frac{k_{\rm NR}}{k}\right)^2\right)\fnu\right].
\end{align}
The normalization factor $\mathcal{N}$ is set by the fact that on very large scales the neutrinos have no effect. Mathematically, this translates to the requirement that, on very large scales, $P(q,s) = P_{\fnu=0}(q,s)$; from this we conclude that $\mathcal{N}=1$. We then have:
\begin{align}
    \Delta(q\rightarrow\infty)&=\frac{\delta P}{P_{\fnu=0}} = -\frac{38}{25}\fnu+\frac{6}{5} s_{\rm T}\fnu    \label{mainresult}\\
    &= -\left(14.69 + \frac{6}{5}\ln{\left(\Omega_{{\rm m}} h^2 \fnu\right)}\right)\fnu.
    \label{mainresultII}
\end{align}
Although this result is obtained by assuming a matter-dominated Universe, this formula may be used to calculate the power spectrum suppression in a cosmology with dark energy. We note that the $\Omega_{\rm m}$ that appears above is originated from the definition of transition time, (\ref{transitionaT}), and is not related to the dynamical equations. 
\begin{figure}%
    \centering
    \subfloat{\includegraphics[width=10cm]{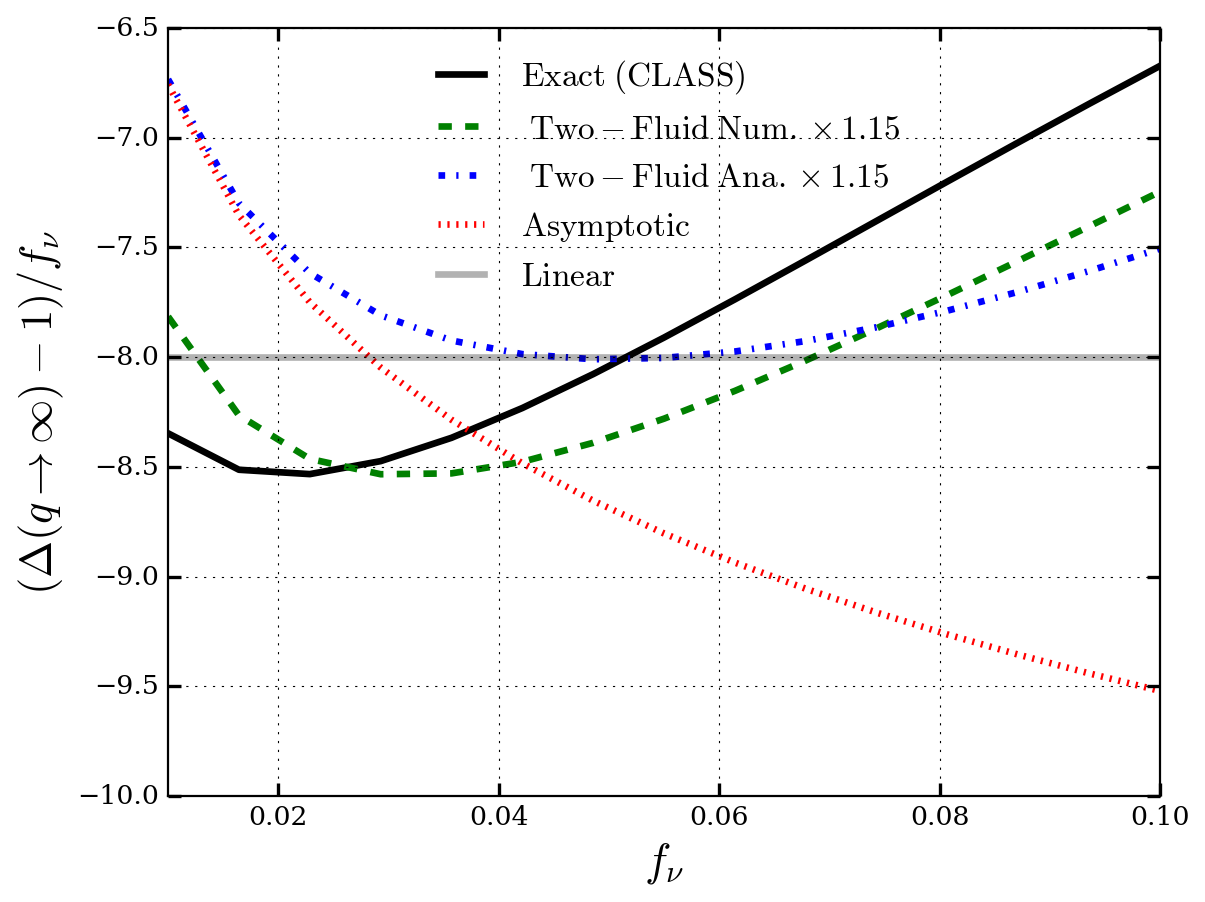} }%
    \caption{The overall suppression of the linear matter power spectrum (\ref{mainresultII}) in the small-scale limit as a function of $f_{\nu}$. The solid gray horizontal line is 
    the linear prediction of \citep{Hu_1998}, $1 - 8 \fnu$, which leads to $\Delta - 1)/\fnu$ of $-8$. The short-dashed red curve is our asymptotic formula (\ref{mainresultII}), with $\Omega_{\rm m}=0.315$. The long-dashed green curve and the solid black curve give, respectively, results from numerical solution of the two-fluid system (\ref{dcb})-(\ref{tnu}) (where we consider a $\Lambda$CDM universe instead of an EdS one), and from \textsc{class}. The dash-dotted blue curve shows our analytic  closed-form formula (\ref{closed_form}). The agreement between the two-fluid scheme and \textsc{class} improves with increasing neutrino mass. We multiply the two-fluid predictions by 1.15 for better visualization.}%
    \label{Delta_minus_one_over_f}%
\end{figure}
We show the power spectrum suppression on small scales, defined as $\Delta(q\rightarrow \infty)$, as a function of the neutrino mass fraction, $f_{\nu}$ in Fig. \ref{Delta_minus_one_over_f}. This figure demonstrates that the analytic  approach in (\ref{closed_form}) (dotted-blue curve) gives a very good approximation to the result from numerically solving the full two-fluid system (dashed-green curve). The full numerical solution is obtained by including the dark energy in the Hubble parameter ($\mathcal{H(\tau)}$), gravitational potential ($\phi(\tau)$) and matter density ($\Omega_{\rm m}(\tau)$) and numerically solving the coupled-differential equations, (\ref{dcb})-(\ref{tnu}). We also use the exact free-streaming scale (\ref{kfree}) instead of the approximation we made in (\ref{kfree0}) and (\ref{kNR}). The asymptotic formula (\ref{mainresultII}) is not accurate if the neutrino mass becomes large. However, it may be used in place of the constant suppression if needed.

\section{Consistency of Fisher Forecasts}
\label{Fisher Forecast}

A Fisher forecast shows the expected precision on a given set of parameters by using the derivatives of the model with respect to them, weighted by the inverse covariance \cite{Tegmark_1997}. Here we explore whether our result for the power spectrum can replicate the exact Fisher matrix from \textsc{class}. We start by assuming a survey with similar effective volume to BOSS DR12 CMASS sample \cite{dawson2012baryon,Anderson_2014,reid2016sdss}, $V_{\rm eff}=3\,{h^{-3}\rm Gpc^3}$, in a cubic box of size $L \equiv V^{1/3}$, centered at the mean redshift of $\bar{z}=0.57$. The wave-number corresponding to this volume is the fundamental mode $k_{\rm F} = 2\pi/L$. We need to bin the Fourier modes into evenly spaced intervals of $\Delta k$. In our forecast, we take $\Delta k = 0.005\, h{\rm Mpc}^{-1}$. The forecast will be presented for the redshift-space galaxy power spectrum monopole including the Kaiser factor as \cite{1987MNRAS.227....1K}:
\begin{align}
    P_{\rm gg}(k) =\frac{1}{2} \int_{-1}^{1}b_1^2(1+\beta \mu^2)^2 P(k) d\mu = P(k)b_1^2\left(1+\frac{2\beta}{3}+\frac{\beta^2}{5} \right)
    \label{GalaxyPowerSpec}
\end{align}
where $\beta = f/b_1$ and $f\equiv \partial \ln{D(a)}/\partial \ln{a}$ is the derivative of the logarithmic growth rate with respect to the logarithmic scale factor. In a $\Lambda$CDM Universe, it is approximated as $f =\Omega_{\rm m}^{0.55}$ \cite{peebles1976peculiar, Bernardeau_2002}. As (\ref{GalaxyPowerSpec}) shows, the amplitude of the clustering, $\sigma_8$, is perfectly degenerate with the linear bias and $\beta$. To address this degeneracy, we introduce a new parameter $\mathcal{A}^2 = \sigma_8^2 b_1^2 \left(1+2\beta/3+\beta^2/5 \right)$ that does not suffer from this issue. Our power spectrum model is then given by $P_{\rm gg} = \mathcal{A}^2 P(k)$.

The Fisher matrix elements may be written as the trace of two power spectrum derivatives (with respect to the parameters), weighted by the inverse covariance matrix:
\begin{align}
    F_{ij} = {\rm Tr}\left[\frac{\partial P_{\rm gg}}{\partial \theta_{i}}\mathbf{C}^{-1}\frac{\partial P_{\rm gg}}{\partial \theta_{j}}\right]
    \label{FisherMatrix}
\end{align}
where $\theta$ is the set of parameters of interest and $\mathbf{C}$ is the covariance matrix of the power spectrum monopole. The Gaussian (linear) contribution to the covariance matrix is:
\begin{align}
    C(k_i, k_j) = \frac{2}{4\pi k^2 \Delta k V_{\rm eff}}\left(P_{\rm gg}(k)+\frac{1}{\bar{n}}\right)^2\delta^{\rm K}_{ij}
\end{align}
where $k_i$ and $k_j$ are the $i^{\rm th}$ and $j^{\rm th}$ bin in Fourier space and $\delta^{\rm K}_{ij}$ is the Kronecker delta. We note that we have used the effective volume rather than the total volume of the survey to obtain a more conservative forecast.

\begin{figure}%
    \centering
    \subfloat{\includegraphics[width=8cm]{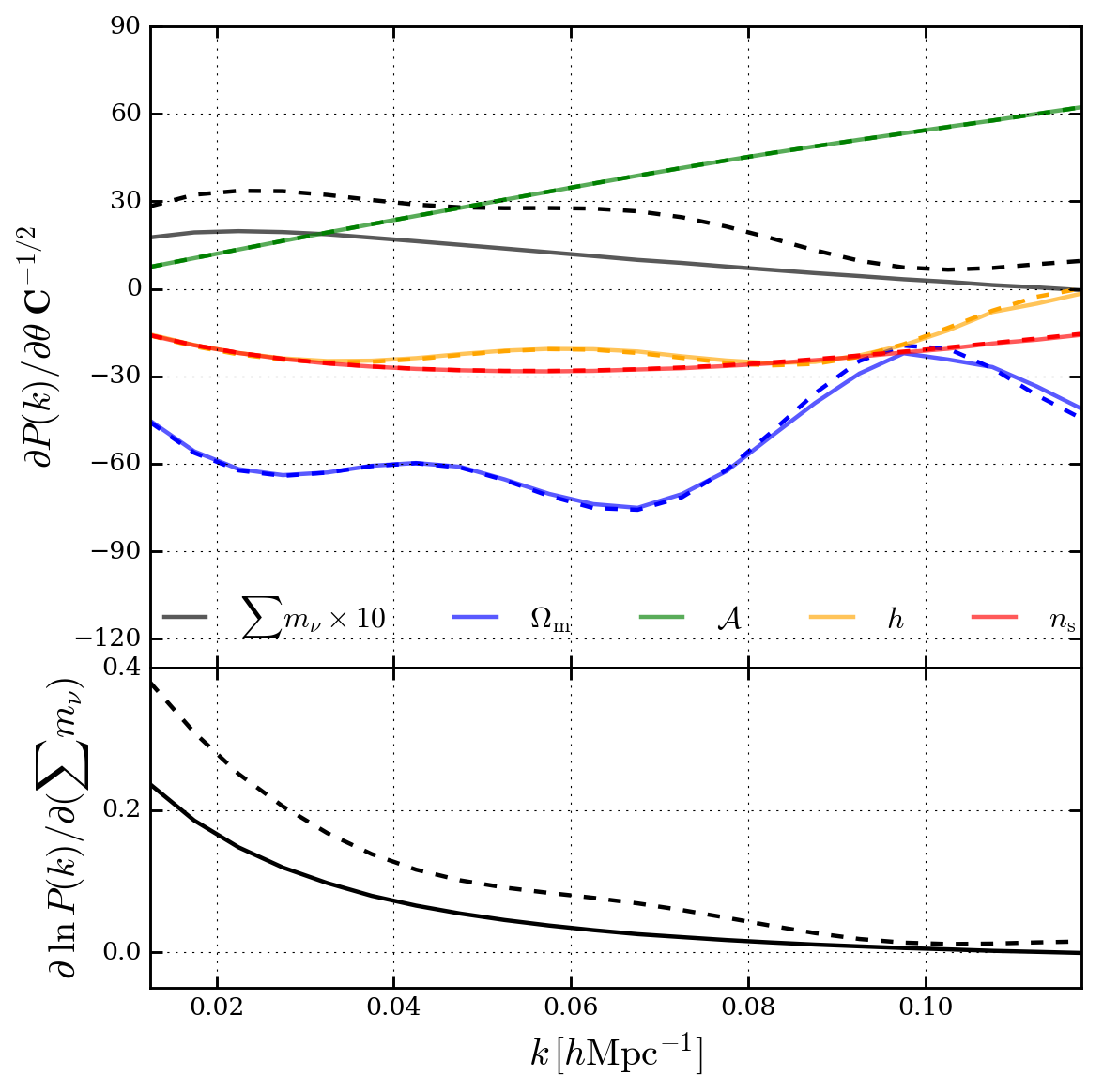} }%
    \caption{\textit{Upper:} The power spectrum monopole response functions defined as $\partial P_{\rm gg}(k)/\partial \theta \mathbf{C}^{-1/2}$ with respect to the parameters of interest, the neutrino mass sum, the matter density, the power spectrum amplitude that we define, $\mathcal{A}$,(\S\ref{Fisher Forecast}, below (\ref{GalaxyPowerSpec})), and the Hubble constant and scalar spectral tilt. The dashed lines are from our two-fluid analytic  model (\ref{closed_form}), while the solid lines are from \textsc{class}. We have multiplied the neutrino curve by 10 to make the vertical axis labels more compact. \textit{Lower:} The derivative of the log power spectrum with respect to the neutrino mass alone, which is essentially what enters the Fisher matrix in (\ref{FisherMatrix}). This panel highlights that our two-fluid model
    produces a derivative with respect to neutrino mass that agrees reasonably well with that from \textsc{class}.}%
    \label{fig:derivatives}%
\end{figure}

We take our analytic  power spectrum ratio written explicitly in (\ref{closed_form}) and multiply it by the power spectrum we get from \textsc{class} with $\sum m_{\nu}=0$. We use the fiducial cosmological parameters: $\theta \equiv \{\sum m_{\nu}=0.131\,{\rm eV},\Omega_{\rm m}=0.315,\Omega_{\rm b}=0.045,h=0.67,n_{\rm s}=0.96, \sigma_8 = 0.834, b_1 = 2.0\}$. The response function for both models, which is the derivative of the galaxy power spectrum with respect to the parameters weighted by the half inverse covariance matrix, is shown in the top panel of Fig. \ref{fig:derivatives}. In the bottom panel, we display the derivative of the logarithmic power spectrum with respect to the neutrino mass. 

As we see from the figures, the main difference between the two models is the neutrino response function. The difference in the neutrino response function is not trivial since our analytic  model also depends on the matter density and Hubble constant as well. However, the two models show good agreement on those parameters. Our analytic  model overestimates, relative to \textsc{class}, The neutrino response function derivatives on large scales. On smaller scales, their agreement improves. The overestimation of the derivatives on the larger scales is not an issue in the Fisher forecast since they are not where most of the constraint is coming from in any case.

\begin{figure}[h]
    \centering
    \subfloat{\includegraphics[width=0.9\textwidth]{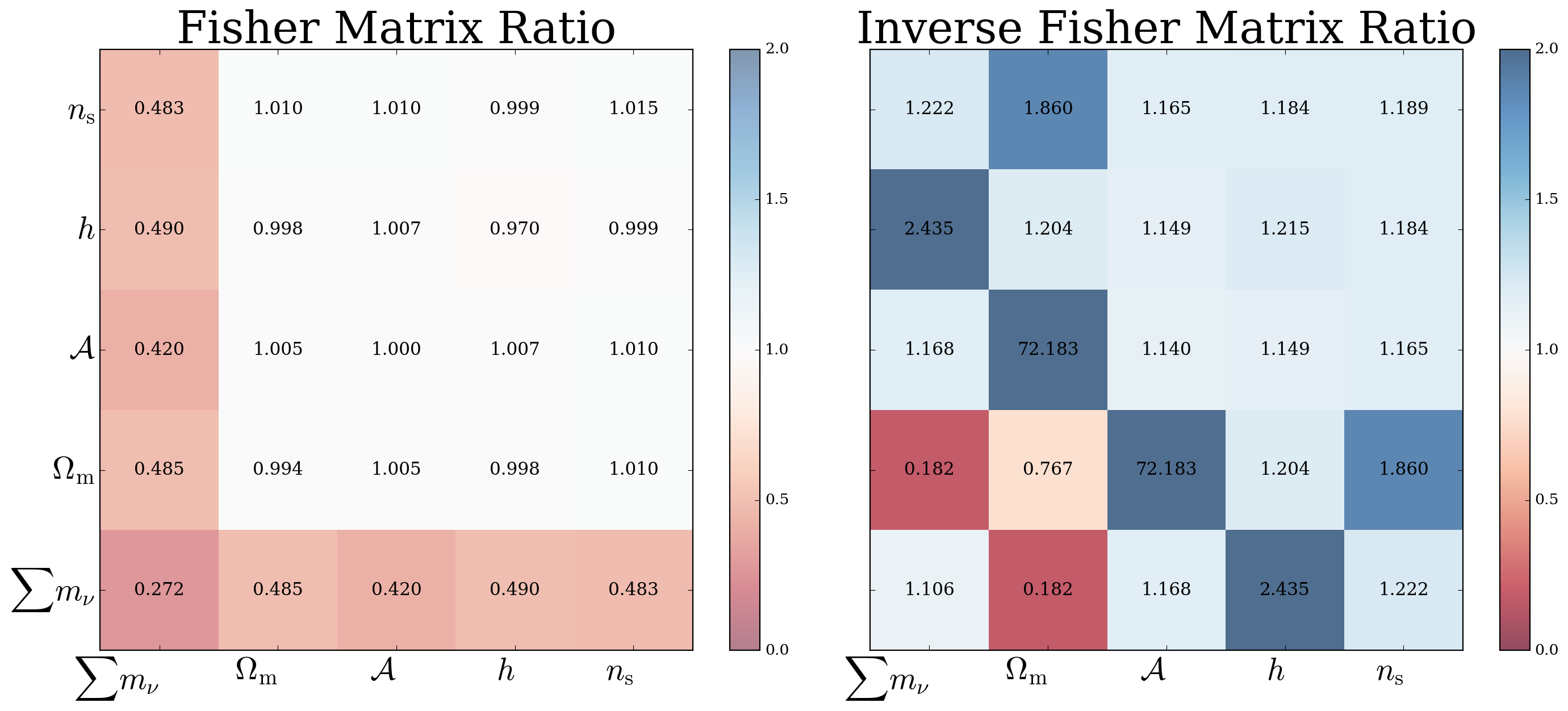} }%
    \caption{\textit{Left:} The ratio of the Fisher matrix elements computed from our two-fluid analytic  model (\ref{closed_form}) to those from \textsc{class} (\ref{FisherMatrix}). As we see, the columns and rows associated with $\sum m_{\nu}$ are the most discrepant from unity (for instance, the $\sum m_{\nu}$ element is $0.272$). \textit{Right:} We show the ratio of the inverse Fisher matrix computed from our two-fluid model to that computed from \textsc{class}. Since inverting a matrix  mixes each element with the others (unless the matrix is diagonal), the localized disagreement in the left-hand panel spreads throughout the inverse Fisher matrix. This will cause slight differences in the forecast between the two approaches, as we further explore in Fig. \ref{fig:Fisher Forecast}.}%
    \label{fig:Propagation}%
\end{figure}

The difference between the \textsc{class} prediction and the two-fluid iterative analytic  solution will distort the forecast to some extent. Fig. \ref{fig:Propagation} shows the ratio between the Fisher information matrices and then the ratio between inverse Fisher matrices in two models and explains how the neutrino mass difference between the two models spreads to other parameters when we invert the matrix. 

\begin{figure}[h]
    \centering
    \subfloat{\includegraphics[width=0.9\textwidth]{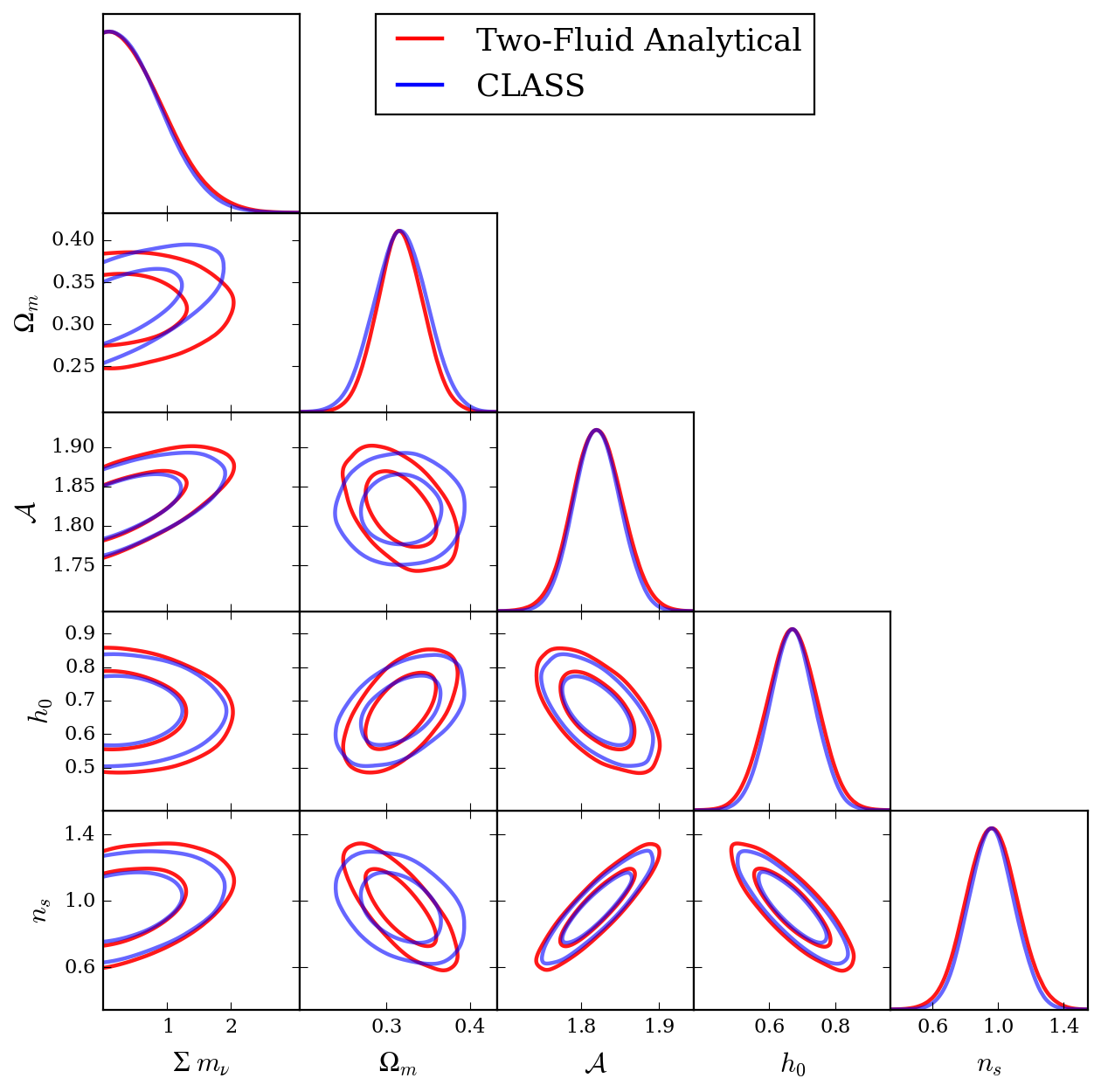} }%
    \caption{Fisher forecast performed with the analytic  model derived in this work (red) and using \textsc{CLASS} (blue). The discrepancies in the $\Omega_{\rm m}$ row and column stem from  the difference in the derivatives as shown in Fig. \ref{fig:derivatives}; this impacts the Fisher matrix elements as we show in Fig. \ref{fig:Propagation}. The forecast is performed at $k_{\rm max} = 0.12\; h{\rm Mpc}^{-1}$. The upper bound on the neutrino mass is $\sum m_{\nu}<1.648 \,{\rm eV}$ at $2\sigma$. Also, as we see, both the error-bars and parameter correlations are very similar between the two models.}%
    \label{fig:Fisher Forecast}%
\end{figure}

Finally, in Fig. \ref{fig:Fisher Forecast} we present the Fisher forecast comparison between the two models. In this forecast we use $k$  from $k_{\rm min}=0.01\,h{\rm Mpc}^{-1}$ to $k_{\rm max}=0.12\,h{\rm Mpc}^{-1}$. The ellipses show the $68\%$ and $95\%$ confidence levels. This figure shows that using our (\ref{closed_form}) formula can re-produce the same forecast as \textsc{class}. The differences between the forecasts are due to the fact that changing the cosmological parameters ($\Omega_{\rm m}, h$) will change the $f_{\nu}$ according to (\ref{fnu0}). This means that we are using (\ref{closed_form}) for a different effective neutrino mass and therefore the predictions would be different. As seen from Fig. \ref{fig:derivatives}, the derivative of power spectrum with respect to parameters are slightly different between the two models. This will give rise to the discrepancy we see in Fig. \ref{fig:Propagation}, which is why the forecasts are slightly different. We also recall that we have used the BOSS DR12 CMASS volume and mean number density, of Luminous Red Galaxies (LRGs).

The error-bars on the parameters from the two models overall appear to be very similar. The parameter correlations also seem to agree between the two models. For the two-fluid iterative model, the $1\sigma$ error-bar on the neutrino mass is $\sigma_{\sum m_{\nu}}=0.763\,{\rm eV}$ while for  \textsc{class} it is $\sigma_{\sum m_{\nu}}=0.726\,{\rm eV}$. For other parameters the error-bars are roughly similar and are: $\sigma_{\Omega_{\rm m}}=0.028 \,(0.032)$, $\sigma_{\mathcal{A}} = 0.032 \,(0.032)$, $\sigma_{h}=0.075 \,(0.068)$ and $\sigma_{n_{\rm s}}=0.153\, (0.141)$ for the two-fluid iterative (\textsc{class}) models, respectively.

In Appendix \S\ref{App2} we show that the consistency of our forecast is not a coincident. We repeat the forecast for different neutrino masses and show the marginalized $68\%$ confidence interval in Fig. \ref{forcastformnu}. Our analytic  model can give the exact same Fisher forecast as \textsc{class} for $\sum m_{\nu}<0.25 \;{\rm eV}$. for higher neutrino masses, our model deviates from \textsc{class}. However, as said earlier, the $95\%$ confidence level of the latest neutrino mass measurement from \textit{Planck} \cite{aghanim2020planck} is $0.26\;{\rm eV}$ which indicates that in the mass regions of interest, our model is indistinguishable from \textsc{class}.
\subsubsection*{Error-Bars vs. $k_{\rm max}$}

Adding more modes in the analysis of the power spectrum analysis significantly improves the parameter constraints. Although linear theory should not be relied upon on small scales, studying the error-bars as we go to smaller scales is important because the linear power spectrum monopole does not have the intrinsic complexities of non-linear structure formation such as counter-terms, higher-order biases and redshift-space PT kernels. Also, the error-bars provide an expectation of parameter uncertainties in the quasi-linear regime $(0.1\,h{\rm Mpc}^{-1}<k_{\rm max}<0.15 \,h{\rm Mpc}^{-1})$. 

\begin{figure}[h]
    \centering
    \subfloat{\includegraphics[width=0.6\textwidth]{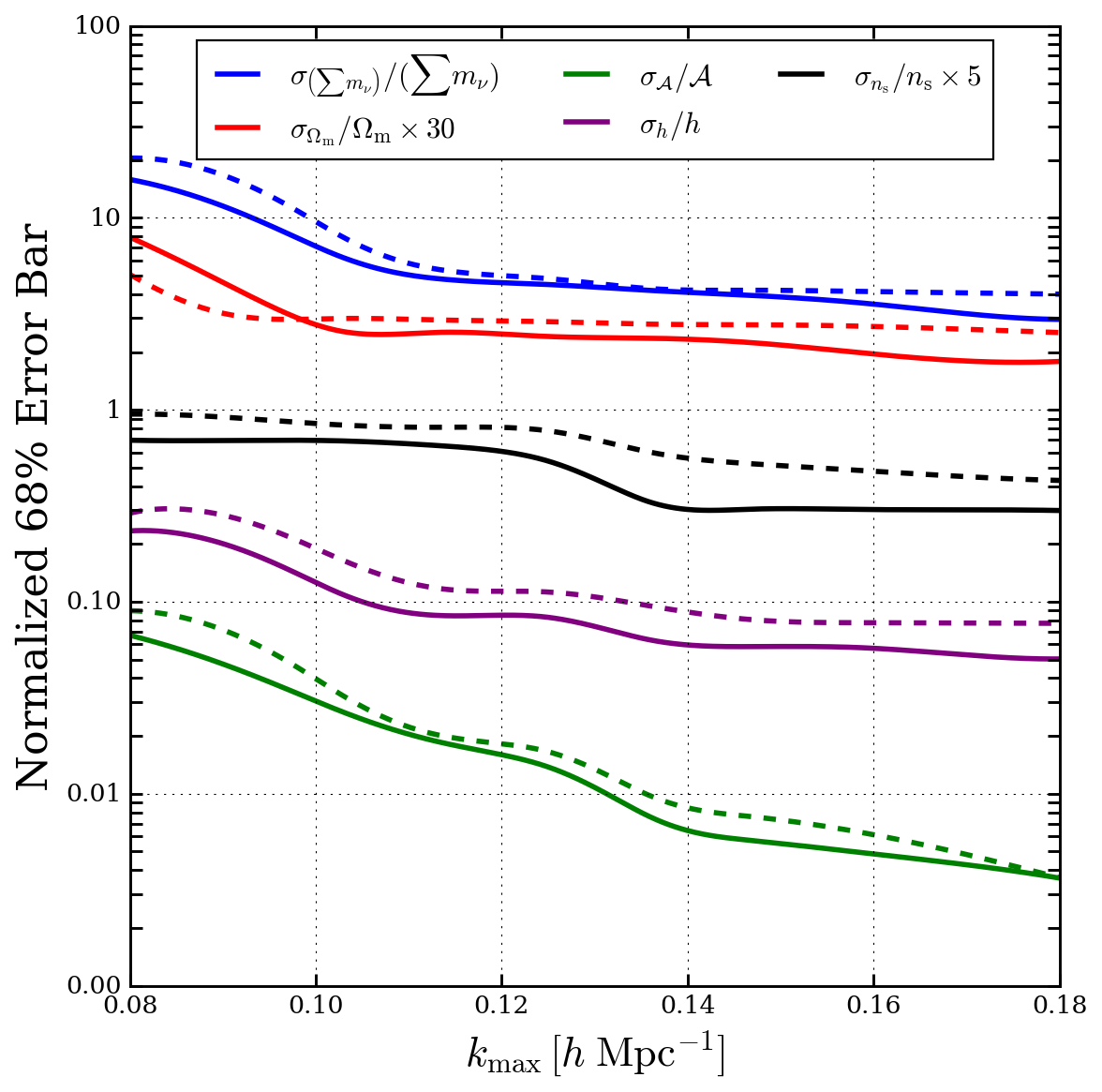} }%
    \caption{The $68\%$ errorbar on the parameters of interest vs. $k_{\rm max}$. As seen from the plot, the parameter constraints get better as we go to higher $k_{\rm max}$, except on $\sum m_{\nu}$ and $\Omega_{\rm m}$, which saturates at $k_{\rm max}\approx0.11\, h{\rm Mpc^{-1}}$. As explained in the main text, the galaxy power spectrum does not gain more constraining power as we go to smaller scales.}%
    \label{fig:kmax}%
\end{figure}

We plot the error-bar for each parameter vs. $k_{\rm max}$ in Fig. \ref{fig:kmax}. As we see, for all parameters, the error-bar improves significantly as we go to higher $k_{\rm max}$. However, it seems that the power spectrum is saturated for both the neutrino mass and the matter density at $k_{\rm max}>0.1\,h{\rm Mpc}^{-1}$. For the matter density, the precision we find is about $11\%$ for  $k_{\rm max}\approx0.1\,h{\rm Mpc}^{-1}$.

The $95\%$ confidence level (CL) upper bound for the neutrino mass on linear scales is $\sum m_{\nu}<1.648 \,{\rm eV}$ which is much less constraining relative to \textit{Planck}+BAO. \cite{Ivanov_2020} finds that the upper bounds on the neutrino mass sum using a full-perturbative model up to $k_{\rm max} = 0.3\;h{\rm Mpc}^{-1}$ with the Gaussian priors from the final \textit{Planck} data release is $\sum m_{\nu}<0.84\,{\rm eV}$ with $95\%$ CL. This result suggests that the latest BOSS data release can only be used to break the degeneracies within the \textit{Planck} data. Further, \cite{Ivanov_2020} notes that the BOSS data alone can only rule out very large neutrino masses, $\sum m_{\nu} > 1\;{\rm eV}$. We note that we have not used any prior on the parameters in contrary to \cite{Ivanov_2020}. We only fix the $\Omega_{\rm b}$ due to lack of constraining power in measurement of the baryonic density from LSS. \cite{noriega2024unveiling} measures the neutrino mass on the BOSS DR12+eBOSS datasets with the same prior scheme as our forecast. Their measurement on the BOSS DR12 is very close to our forecast. These full-shape power spectrum measurements and our forecast strongly suggest that the power spectrum alone does not have enough constraining power for the neutrino mass with the current survey volume. Increasing the $k_{\rm max}$ adds more information but we also have more parameters to infer such as non-linear biases and EFT counter-terms \cite{leonardo2017effective}. On the other hand, the power spectrum reaches a plateau for higher $k$'s (for $k>k_{\rm FS}, P\rightarrow P_{f_{\nu}=0}(1-8f_{\nu})$) which indicates that the neutrino mass bounds do not change as we go to smaller scales. In the simplest possible case, we may calculate the Fisher matrix for only the neutrino mass and analytically prove that $\sigma_{\sum m_{\nu}}/(\sum m_{\nu})\propto1/\sqrt{k_{\rm max}}$. All in all, we interpret this forecast as a need for higher-order correlation statistics for the neutrino mass measurements such as proposed in \cite{kamalinejad2020non, Hahn_2021, ruggeri2018demnuni}.
\section{Concluding Remarks}

In this paper, we have solved the two-fluid equations for the cold dark matter+baryons and massive neutrinos iteratively in $f_{\nu}$ and found a closed-form formula for the power spectrum suppression due to neutrino free-streaming. We built our model on a simple and physically-motivated model for the equation of state of neutrinos, that is in good agreement with the exact results form \textsc{class}. Our power spectrum model (\ref{closed_form}) is able to replicate the exact predictions from \textsc{class} to better than 5\% on scales smaller than $0.1\;h{\rm Mpc}^{-1}$. Our model shows that complicated tracing of the higher moments ($\ell's$) in the Boltzmann hierarchy is not necessary.

This closed-form formula may accelerate the total matter power spectrum calculation since we do not need to run \textsc{class} to get the power spectrum. This formula is presented in (\ref{closed_form}) and potentially accelerates the computation.

Our approach also may be used to calculate an approximation for the cross-power spectrum between cold dark matter+baryons and neutrinos, which is needed for the higher-order corrections in the matter power spectrum and bispectrum \cite{ruggeri2018demnuni, hahn2020constraining, hahn2021constraining, kamalinejad2020non}.

Another use of this formula is in forecasting for the power spectrum and bispectrum. In this paper, we have produced a Fisher forecast for both the \textsc{class} and analytic  power spectra and found that the error-bars are nearly identical between the two. The $1\sigma$ error-bar on the neutrino mass given by the forecast is $\sigma_{\sum m_{\nu}}\simeq 0.75 \,{\rm eV}$ and is consistent with the previous work with the same survey volume \cite{Ivanov_2020, noriega2024unveiling}.




\label{lastpage}

\acknowledgments

The authors thank R. Cahn, J. Chellino, A. Greco, J. Hou, A. Krolewski, W. Ortolà-Leonard, O. Philcox, and M. Reinhard for useful discussions.
The authors also thank A. Greco and J. Hou for helping make the plots. We are also thankful to other members of Slepian research group for useful insights and discussions.


\bibliographystyle{plain}
\bibliography{Ref.bib}
\appendix

\section{Dark Energy Effects}
\label{Dark Energy Modifications}

In this appendix, we obtain the asymptotic formula for the power spectrum suppression on small scales, $k\gg k_0$, similar to what we did in section \S\ref{limiting form}.

Let us imagine a gravitational potential well with some arbitrary depth along the line of sight. As the photons enter the well, they gain energy, and as they leave the well, they lose it. If the potential well does not change as they traverse it, the photons' net energy gain is zero (exactly what happens in an Einstein-de Sitter universe). However, when DE dominates, the potential changes as the Universe expands; therefore the photons will experience gravitational redshift. This is the late-time Integrated Sachs Wolfe
(ISW) effect in the CMB \cite{Sachs:1967er}. Since in the DE-dominated Universe, the potential is not constant, its rate of change will affect the clustering of matter and of neutrinos on all scales. 

We incorporate DE in our fluid equations as follows. on scales smaller than $k_0$, the effect of the time-varying gravitational potential will be eliminated since it is coupled to the matter perturbations via the Poisson equation (\ref{Poisson}) and hence suppressed by a factor of $1/k^2$. We may rewrite the time derivative of the potential as:
\begin{align}
    \phi(k, a) = \frac{3}{2}\Omega_{\rm m}\left(\frac{a_0^2 H_0^2}{k^2}\right)\left(\frac{\delta}{a}\right) \propto \left(\frac{\text{Scales of Interest}}{\text{Horizon Size}}\right)^2
\end{align}
Since the scales of interest are smaller than the free-streaming scale of neutrinos, we may neglect the effects of the gravitational potential's evolution. 

Second, we implement the iterative method we discussed earlier. While we ignore the evolution of $\phi$, the DE changes the rate at which the Universe is expanding. We perturbatively expand in $\fnu$ and order by order, solve the fluid equations. The system of equations we need to solve is exactly that of (\ref{dcb''}) and (\ref{dnu''}) but  with 
\begin{align}
    \frac{\mathcal{H}'}{\mathcal{H}}(s)+1 = 2- \frac{3}{2}\left(\frac{1}{1+\Omega_{\Lambda}/\Omega_{\rm m}a^3}\right)= 2-\frac{3}{2}\left(\frac{1}{1+e^{3(s-s_{\rm{eq}})}}\right),
    \label{H}
\end{align}
where $s_{\rm{eq}}$ is the value of $s$ at matter-dark energy equality. As we see, at early times, before equality, $s\ll s_{\rm{eq}}$ and therefore (\ref{H}) converges to $1/2$, recovering the matter-dominated solutions. The growing solution of the zeroth order CDM+baryon equation (\ref{dcb''}) then takes the form:
\begin{align}
    \dcb^{0}(q,s) = \mathcal{C}\,\dcb(k,s)\, \fone\left(\frac{1}{3},1,\frac{11}{6},-e^{3\left(s-s_{\rm{eq}}\right)}\right)
    \label{lambda matter solution}
\end{align}
where $\mathcal{C}$ is a normalization factor, $\dcb(k,s)$ is the present-time matter perturbation in an EdS universe and $\fone$ is the Gauss hypergeometric function. (\ref{lambda matter solution}) is consistent with earlier works on the effect of dark energy on the growth of matter perturbations such as \cite{linder2003cosmic}.

\begin{figure}%
    \centering
    \subfloat{\includegraphics[width=9cm]{deltamatter0growth.pdf} }%
    \caption{The matter growth function vs. scale factor for two different cosmologies, the first (dashed black) with $\Omega_{\Lambda} = 0$, and the second (solid green) with  $\Omega_{\Lambda} = 0.7$. We see the suppression of growth in the presence of dark energy. $a_0=1$ is the present-time scale factor.}%
    \label{deltagrowth}%
\end{figure}

Fig. \ref{deltagrowth} shows how dark-energy domination changes the growth of the matter perturbations. It essentially suppresses the growth factor by almost $20\%$ at the present time. From the previous section it is clear that on small scales, the neutrinos are almost unperturbed. Therefore, we are safe to assume that on these scales $\dnu(q\rightarrow\infty,s) \approx 0$. 
The solution scheme in this case, as is generic for a second-order differential equation, is:
\begin{align}
    \dcb^{(1)}(q,s) = G(s)\left(c_1+Y_{{\rm P},G}\right)+D(s)\left(c_2+Y_{{\rm P},D}\right).
\end{align}
$G(s)$ is the growing mode and $D(s)$ the decaying mode of the homogeneous equation, $c_1$ and $c_2$ are constants, and $Y_{{\rm P},G}$ and $Y_{{\rm P},D}$ are the particular solutions of the non-homogeneous differential equation with a driving force of the form (\ref{lambda matter solution}). At the end, we need to eliminate the decaying mode since at early times it diverges. Therefore, we tune $c_2$ to remove it. This is only possible in a universe where DE has taken over the expansion relatively recently. If $s_{\rm{eq}}$ is large enough, we cannot perform this tuning since the particular solution will not be close enough to a constant.

Therefore, the matter perturbation up to first order becomes:
\begin{align}
    \dm(q,s)&=\mathcal{C}\,\dcb(q,0)e^s\, \fone\left(\frac{1}{3},1,\frac{11}{6},-e^{3\left(s-s_{\rm{eq}}\right)}\right)
    \label{first order matter pt}\\&\times\left(1-\fnu+ \fnu \frac{\Gamma \left(\frac{11}{6}\right) \left[G_{3,3}^{2,2}\left(e^{3 s-3s_{\rm{eq}}} \bigg|
\begin{array}{c}
 0,\frac{2}{3},1 \\
 0,0,-\frac{5}{6} \\
\end{array}
\right)-G_{3,3}^{2,2}\left(e^{3s_{\rm T}-3 s_{\rm{eq}}} \bigg|
\begin{array}{c}
 0,\frac{2}{3},1 \\
 0,0,-\frac{5}{6} \\
\end{array}
\right)\right]}{5\,\Gamma \left(\frac{1}{3}\right)}\right),\nonumber
\end{align}
where $G_{3,3}^{2,2}$, is the Meijer $G$-function. One important consistency check is to verify that (\ref{first order matter pt}) converges to (\ref{dm}) when $s_{\rm{eq}}\rightarrow0$. In this limit, the two solutions do become the same up to zeroth order in $1/q^2$. With the consistency check passed, we find the power spectrum suppression as:
\begin{align}
    \frac{\delta P}{P_{\fnu=0}}=-2\fnu\Bigg(1+\frac{\Gamma \left(\frac{11}{6}\right) \left[G_{3,3}^{2,2}\left(e^{3 s-3s_{\rm{eq}}}\bigg|
\begin{array}{c}
 0,\frac{2}{3},1 \\
 0,0,-\frac{5}{6} \\
\end{array}
\right)-G_{3,3}^{2,2}\left(e^{3s_{\rm T}-3 s_{\rm{eq}}} \bigg|
\begin{array}{c}
 0,\frac{2}{3},1 \\
 0,0,-\frac{5}{6} \\
\end{array}
\right)\right]}{5\,\Gamma \left(\frac{1}{3}\right)}\Bigg).
\label{DEPSsupression}
\end{align}

\section{Forecast Results for Different $\sum m_{\nu}$}
\label{App2}

In the main text, we presented our Fisher forecast for a fiducial neutrino mass of $\sum m_{\nu} = 0.131\;{\rm eV}$, corresponding to $f_{\nu} = 0.01$. It is natural to ask how the agreement of our approach with \textsc{class} changes if we change the neutrino mass.
To see whether our model in (\ref{closed_form}) gives acceptable results if we change the fiducial value of $\sum m_{\nu}$, we perform the forecast for a range of different neutrino masses and compare the $68\%$ CL from our two-fluid model with that of \textsc{class}.
\begin{figure}[h]%
    \centering
    \subfloat{\includegraphics[width=8cm]{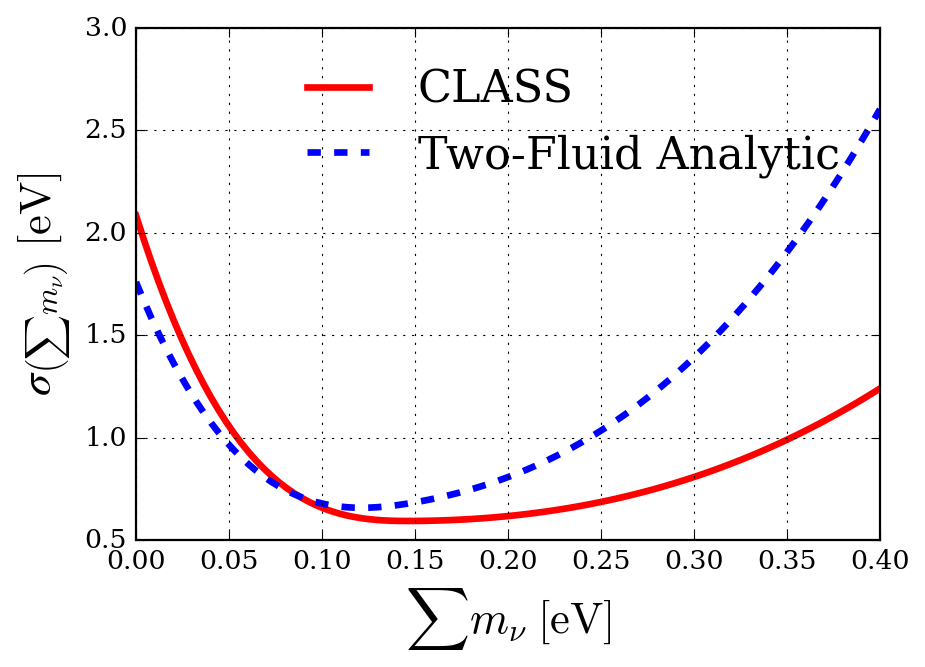} }%
    \caption{Marginalized $1\sigma$ error-bar on $\sum m_{\nu}$ obtained from a similar Fisher forecast to that presented in \S\ref{Fisher Forecast} for a range of neutrino masses. Our analytic two-fluid model (\ref{closed_form}) gives the exact same forecast as \textsc{class} for $\sum m_{\nu}<0.25\;{\rm eV}$.}%
    \label{forcastformnu}%
\end{figure}
Fig. \ref{forcastformnu} shows the marginalized $68\%$ confidence level for different neutrino masses. We see that for $\sum m_{\nu}<0.25\;{\rm eV}$, which is exactly at the $2\sigma$ limit of \textit{Planck}, we obtain a very similar forecast to that of \textsc{class}.

\section{Empirical Adjustment of the Initialization Time}

Observing that our numerical two-fluid result's shape for $P(k)$'s asymptotic behavior in $\fnu$ appeared very close to the result from \textsc{class} motivated us in the main text to explore applying an overall empirical rescaling factor, of $1.15$. This gave very good agreement. In the course of that exploration, we also observed that instead rescaling the initial condition scale factor (which we took to be the transition scale factor $a_{\rm T}$) by $1/3$ also gave very good agreement of our numerical two-fluid approach with the results from \textsc{class}. The effect of this change is initialize our system's evolution earlier. 

We note that this approach is also mathematically equivalent to solving the two-fluid system for a single neutrino mass state rather than for three degenerate states. We note that this re-scaling should not affect the free-streaming scale (\ref{kfree}), its present day value (\ref{kfree0}), or the non-relativistic scale (\ref{kNR}). 
\begin{figure}%
    \centering
    \subfloat{\includegraphics[width=16cm]{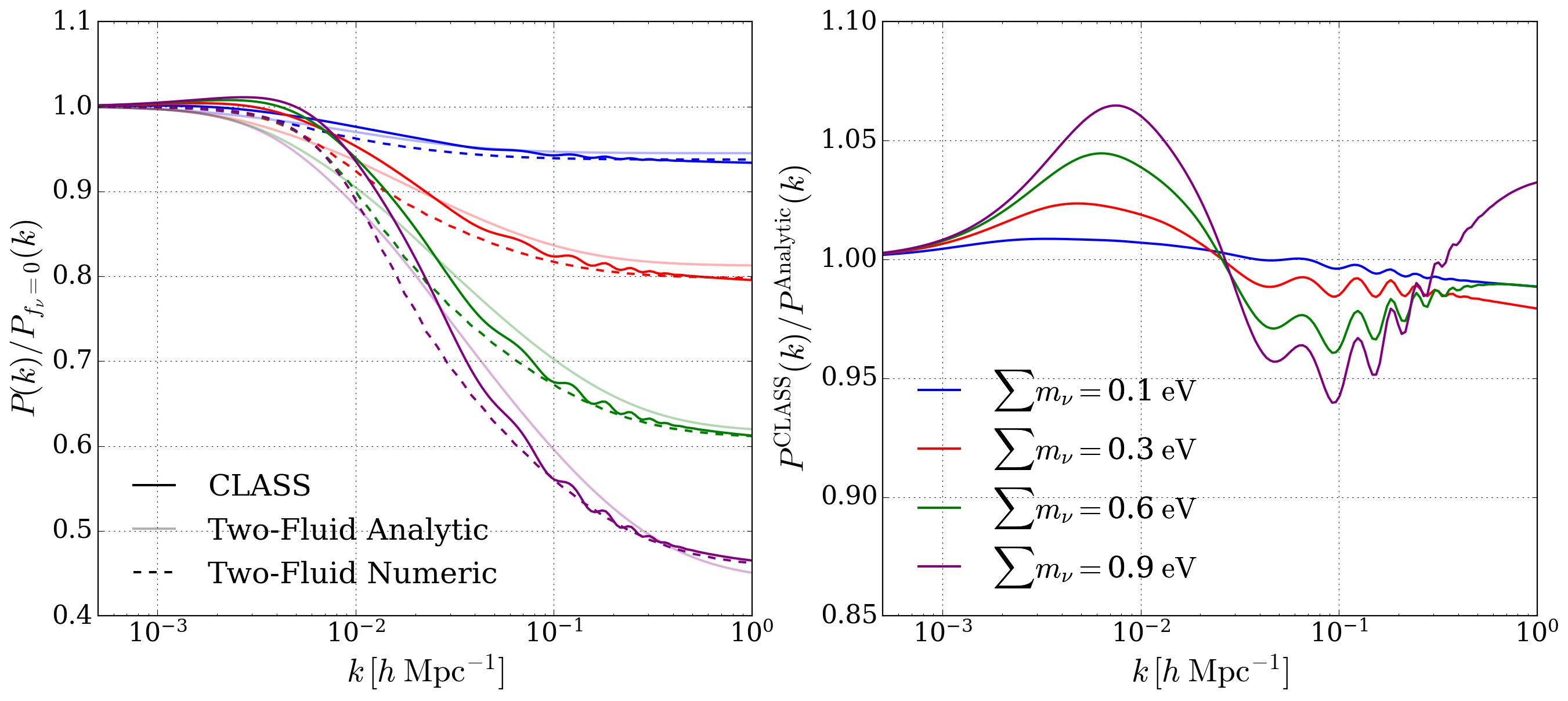} }%
    \caption{These panels are the same as those in Fig. \ref{fig:All three together} except that here we initialize the evolution at $a_{\rm T}/3$, as  discussed in Appendix \ref{App2}. This choice seems to capture the fact that the transition is not completely sharp. Compared to Fig. \ref{fig:All three together}, the agreement with \textsc{class} is very slightly worse here for low $k \lesssim 10^{-2}\;h {\rm Mpc}^{-1}$, but notably better for higher $k \gtrsim 10^{-1}h\;{\rm Mpc}^{-1}$.}%
    \label{wrongplot1}%
\end{figure}


It seems physically correct that increasing the redshift at which we initialize our system does improve the agreement. Our two-fluid approach in this work does not account for higher moments of the Boltzmann hierarchy, such as anisotropic stress, and the neutrinos certainly have such moments, especially before they become non-relativistic \cite{1995, shoji2010massive}. The effect of these higher moments would be to cause power to leak away from the two lowest moments we track here, essentially, reducing the energy available to enable the neutrinos to travel in straight paths outward from their starting points. 
Future work might fruitfully explore whether this empirical rescaling factor can be rigorously derived from the considerations above. In the present work, we simply highlight the remarkable agreement even our simple two-fluid approach (with this one additional free parameter) can obtain with \textsc{class}.

\begin{figure}%
    \centering
    \subfloat{\includegraphics[width=10cm]{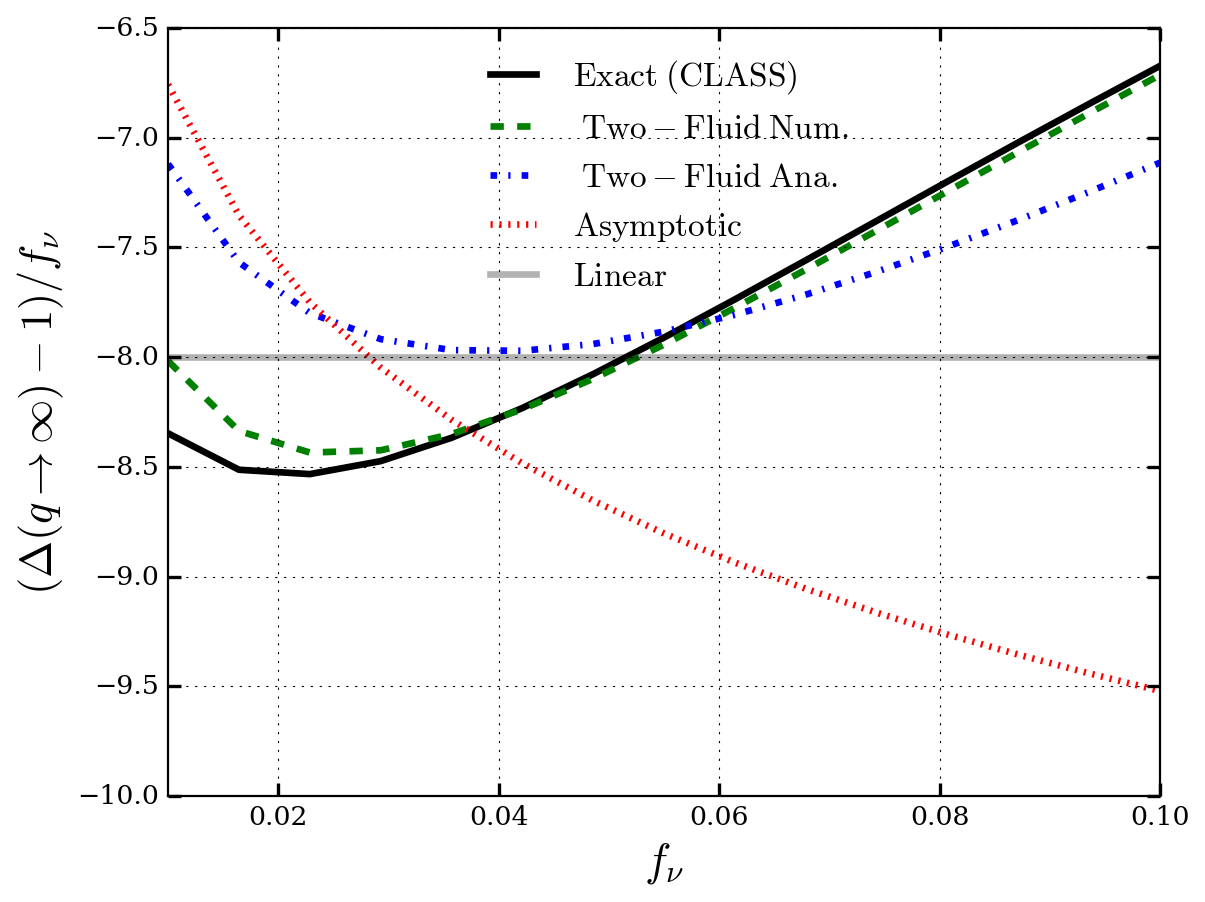} }%
    \caption{The power spectrum suppression $\Delta$ defined in (\ref{mainresult}), vs. the neutrino mass fraction $f_{\nu}$. As in Fig. \ref{wrongplot1}, we here have evolved the system starting at $a_{\rm T}/3$ rather than at $a_{\rm T}$. The two-fluid numerical solution gives impressive agreement with \textsc{class}. The performance of our analytic model (\ref{closed_form}) with the earlier initialization shown here is notably better at higher neutrino masses than the result in the main text (initializing at $a_{\rm T}$), as we see by comparing to Fig. \ref{Delta_minus_one_over_f}.}%
    \label{wrongplot2}%
\end{figure}
\end{document}